\pdfoutput=1
\documentclass[11pt]{article}

\usepackage{graphicx}
\usepackage{amsmath}
\usepackage{amssymb}
\usepackage{dcolumn}
\usepackage{bm}
\usepackage{slashed}
\usepackage{color}
\usepackage{authblk}
\usepackage{ulem}
\usepackage{CJK}

\def\l{{\lambda}}
\def\L{{\Lambda}}
\def\d{{\delta}}
\def\D{{\Delta}}
\def\o{{\omega}}

\def\e{{\epsilon}}

\def\a{{\alpha}}
\def\b{{\beta}}

\def\g{{\gamma}}
\def\G{{\Gamma}}
\def\j{{\psi}}
\def\vj{{\varphi}}

\def\p{{\pi}}
\def\P{{\Pi}}
\def\m{{\mu}}
\def\n{{\nu}}
\def\r{{\rho}}
\def\s{{\sigma}}

\def\th{{\theta}}

\def\ps{{\psi}}

\def\x{{\xi}}
\def\P{{\Pi}}

\def\({\left(}
\def\){\right)}
\def\[{\left[}
\def\]{\right]}

\newcommand{\lag}{\langle}
\newcommand{\rag}{\rangle}

\newcommand{\pd}{{\partial}}
\newcommand{\dg}{\dagger}

\newcommand{\tr}{\text{tr}}

\newcommand{\sgn}{\text{sgn}}

\date{\today}

\begin{document}

\begin{CJK}{UTF8}{gbsn}

\title{\bf Medium Correction to Gravitational Form Factors}

\author[1]{{Shu Lin}
\thanks{linshu8@mail.sysu.edu.cn}}
\affil[1]{School of Physics and Astronomy, Sun Yat-Sen University, Zhuhai 519082, China}
\author[1]{{Jiayuan Tian}}

\maketitle


\begin{abstract}
We generalize the gravitational form factor for chiral fermion in vacuum, which reproduces the well-known spin-vorticity coupling. We also calculate radiative correction to the gravitational form factors in quantum electrodynamics plasma. We find two structures in the form factors contributing to the scattering amplitude of fermion in vorticity field, one is from the fermion self-energy correction, pointing to suppression of spin-vorticity coupling in medium; the other structure comes from graviton-fermion vertex correction, which does not adopt potential interpretation, but corresponds to transition matrix element between initial and final states. Both structures contribute to axial chiral vortical effect. The net effect is that radiative correction enhances the axial chiral vortical effect. Our results clarify the relation and difference between spin-vorticity coupling and axial chiral vortical effect from the perspective of form factors. We also discuss the application of the results in quantum chromodynamic plasma, indicating radiative correction might have an appreciable effect in spin polarization effect in heavy ion collisions.
\end{abstract}

\newpage

\section{Introduction}

The experiments of heavy ion collisions in the past few years have found spin polarization in final state particles \cite{STAR:2017ckg}, which confirms early theory prediction based on angular momentum conservation in off-central heavy ion collisions \cite{Liang:2004ph,Liang:2004xn}. In particular, the global polarization of $\L$ hyperon is well described by thermal model based on spin-vorticity coupling \cite{Gao:2007bc,Huang:2011ru,Jiang:2016woz}, which is considered as evidence for creation of rapid spinning quark-gluon plasma, see \cite{Gao:2020vbh,Liu:2020ymh,Becattini:2020ngo,acta_ex,acta_th} for recent reviews. However, prediction for local polarization based on the same picture \cite{Becattini:2017gcx,Wei:2018zfb,Fu:2020oxj} differs almost by a sign from experimental data \cite{STAR:2019erd}, which triggers studies on contributions apart from vorticity. Indeed, recent studies indicate particle spin couples to all types of fluid gradient. This has been confirmed in different approaches \cite{Liu:2021uhn,Becattini:2021suc,Hidaka:2017auj}, providing a novel solution to the problem. Phenomenological studies by different groups point to the same trend as the experimental data, though currently no consensus has been reached on quantitative agreement \cite{Fu:2021pok,Becattini:2021iol,Yi:2021ryh,Fu:2022myl,Wu:2022mkr}. It has been indicated that fluid gradient other than vorticity will lead to redistribution of particles in momentum space, giving rise to extra contribution to spin polarization at the same order in gradient \cite{Lin:2022tma}. It has also been pointed out that different choices of energy-momentum tensor (EMT) can lead to different contributions to spin polarization \cite{Liu:2021nyg}.

This paper discusses another possible correction to spin polarization. It is usually believed that the form of spin-vorticity coupling is fixed: $\D H=-\vec{ S}\cdot\vec{ \o}$, with $\vec{ S}$ and $\vec{ \o}=\frac{1}{2}\nabla\times\vec{ v}$ being spin and vorticity respectively. The coefficient of the coupling is not renormalized by interaction, i.e. there is no anomalous gravitomagnetic moment, which is a manifestation of Einstein equivalence principle \cite{Kobzarev:1962wt,Pagels:1966zza}. However, Lorentz invariance is lost in the presence of a medium and the equivalence principle is expected to fail. Early studies have showed inequality of gravitational mass and inertial mass \cite{Donoghue:1984zs,Donoghue:1984ga} in a medium and thus anomalous gravitomagnetic moment is in principle allowed, i.e. spin-vorticity coupling is not protected in medium. Indeed, recent study have found a negative anomalous gravitomagnetic moment \cite{Buzzegoli:2021jeh} for a massive fermion in medium, indicating medium suppression of spin-vorticity coupling. Interestingly, spin-vorticity coupling is closely related to axial chiral vortical effect (ACVE), as the momentum integration of spin polarization gives the axial current. For massless fermions, the medium correction to chiral vortical conductivity points to an enhanced ACVE instead \cite{Hou:2012xg}. In this paper, we will calculate the medium correction to spin-vorticity coupling for massless fermion. On one hand, this can help us understand the relation between spin-vorticity coupling and ACVE. On the other hand, it also provides a new perspective to phenomenological studies of spin polarization.

The paper is organized as follows: in Section~\ref{sec_FF}, we will study scattering of fermion in background metric perturbation and establish a connection between gravitational form factor (GFF) with spin-vorticity coupling; in Section~\ref{sec_HTL}, we will calculate one-loop radiative correction to the GFF. In particular, we will keep medium dependent contribution in the hard thermal loop (HTL) approximation. We will find medium correction to scattering amplitude contains two structures: one structure comes from fermion self-energy, which points to medium suppression of spin-vorticity coupling; the other structure comes from correction to graviton-fermion vertex. It does not adopt potential type of interpretation but corresponds to transition matrix element between initial and final states. Both structures contribute to ACVE. This clarifies the relation between spin-vorticity coupling and ACVE from the perspective of form factors. Our results have infrared divergence when the momentum exchange tends to zero. We obtain infrared safe results after screening effect is taken into account. We find radiative correction leads to enhancement of ACVE; Section~\ref{sec_outlook} is devoted to summary and outlook.

\section{Gravitational form factor and spin-vorticity coupling}\label{sec_FF}

We consider scattering of fermion in background metric field. The interaction vertex of fermion and background metric field can be described by the GFF. For massive fermion, the gravitational form factor is defined as \cite{Polyakov:2018zvc}
\begin{align}\label{gravFF}
  \lag P_2|T^{\m\n}(Q)|P_1\rag=\bar{u}(P_2)\bigg[A(Q^2)\frac{P^\m P^\n}{m}+B(Q^2)\frac{iP^{\{\m}\s^{\n\}\r}Q_\r}{m}+D(Q^2)\frac{Q^\m Q^\n-g^{\m\n}Q^2}{4m}\bigg]u(P_1),
\end{align}
where $P_1$ and $P_2$ are momenta of incoming and outgoing particles (without loss of generality, we discuss scattering of particles). $u(P_1)$ and $\bar{u}(P_2)$ are corresponding wave functions. $P$ and $Q$ are defined respectively as $P=\frac{1}{2}(P_1+P_2)$, $Q=P_2-P_1$. The symmetrization is defined as $a^{\{\m}b^{\n\}}=\frac{1}{2}\(a^\m b^\n+a^\n b^\m\)$. Among the three form factors, $A$ and $B$ describe coupling of particle mass and spin to metric. $D$ exists only for composite particles. We focus on elementary particles below so we ignore $D$.

Note that massless fermion does not have a mass scale, so the above definition does not apply. We need to introduce a new definition
\begin{align}\label{gravFF0}
  \lag P_2|T^{\m\n}(Q)|P_1\rag=\bar{u}(P_2)\bigg[A(Q^2)\frac{P^\m P^\n}{P\cdot n}\pm B(Q^2)\frac{-iP^{\{\m}\e^{\n\}\l\s\r}\g_\l n_\s Q_\r}{P\cdot n}\bigg]u(P_1).
\end{align}
In this definition, $\pm$ corresponds to right/left-handed fermion respectively. $A$ and $B$ are form factors. We have introduced a time-like frame vector $n$, which is similar to the frame vector in chiral kinetic theory \cite{Chen:2014cla,Hidaka:2016yjf}. Although both structures depend on $n$, their sum corresponding to EMT does not. At tree level, $T^{\m\n}=\frac{i}{2}\bar{\j}\(\g^{\{\m}\pd^{\n\}}-\g^{\{\m}\overleftarrow{\pd}^{\n\}}\)\j$. $A$ and $B$ can be fixed by matching two sides of \eqref{gravFF0}. To be specific, we restrict to right-handed fermions in the discussion below. We are interested in the limit $Q\to0$, i.e. a slow-varying metric field. By expanding in $Q$, we can easily fix the form factors. Using $i\pd^\m\to P_1^\m$, $-i\overleftarrow{\pd}^\m\to P_2^\m$, we obtain at $O(q^0)$
\begin{align}\label{fix_A}
  \bar{u}(P)\g^{\{\m}P^{\n\}}u(P)=\bar{u}(P)A\frac{P^{\m}P^{\n}}{P\cdot n}u(P).
\end{align}
We take Weyl representation of gamma matrices, for which we have $\g^\m\to\s^\m$ for right-handed fermion and $n^\m=(1,0,0,0)$. From \eqref{fix_A} we obtain $A=1$. To fix $B$, we need to expand \eqref{gravFF0} to $O(q)$. We use the following explicit wave functions to do the expansion
\begin{align}\label{u12}
  &\bar{u}(P_2)=\frac{1}{\sqrt{2p_2}}\(2p_2\x_2^\dg,0\),\nonumber\\
  &u(P_1)=\frac{1}{\sqrt{2p_1}}\(0,2p_1\x_1\)^T,
\end{align}
with $p_{1,2}$ corresponding to norms of 3-momenta, and $\x_{1,2}$ being $2\times1$ matrices. From \eqref{u12} we obtain the following relations \cite{Dong:2021fxn}
\begin{align}\label{usigma}
  &\bar{u}(P_2)u(P_1)=\x_2^\dg\x_1(4p_1p_2)^{1/2},\nonumber\\
  &\bar{u}(P_2)\s^iu(P_1)=\x_2^\dg\x_1(4p_1p_2)^{1/2}\frac{p_1p_{2i}+p_{1i}p_2-i\e^{ijk}p_{2j}p_{1k}}{p_1p_2+\vec{p}_1\cdot\vec{p}_2}.
\end{align}
We will not expand the common $\x_2^\dg\x_1(4p_1p_2)^{1/2}$ in \eqref{gravFF0}. Using $p_1p_2+\vec{p}_1\cdot\vec{p}_2=2p^2+O(q^2)$ and $p_1p_{2i}+p_{1i}p_2=2pp_i+O(q^2)$, we note that the only $O(q)$ term on the left hand side (LHS) of \eqref{gravFF0} comes from $\e^{ijk}p_{2j}p_{1k}=-\e^{ijk}p_{j}q_{k}$. While the $O(q)$ term on the right hand side (RHS) of \eqref{gravFF0} comes from coefficient $B$. When taking $\m\n=0i$ and $\m\n=ij$, we can fix $B=-\frac{1}{2}$.

Below we show the $B$ term can give the correct spin-vorticity coupling. We consider fluid in equilibrium, and take the frame vector $n^\m$ to be the same as the fluid rest frame vector $u^\m$, i.e. $n^\m=u^\m=(1,0,0,0)$. We introduce specific metric perturbation $h_{0i}(t,x)=v_i(t,x)$, which leads to an effective fluid vorticity
\begin{align}\label{vorticity}
  \o^\m=\frac{1}{2}\e^{\m\n\r\s}u_\n\nabla_\r u_\s\;\to\;\o^i=-\frac{1}{2}\e^{ijk}\pd_j v_k+O(v^2),
\end{align}
with $v_i$ identified as fluid velocity. By choosing proper metric perturbation, we can model arbitrary fluid vorticity. Note that the EMT and the metric field couples as $\frac{1}{2}T^{\m\n}h_{\m\n}$, the scattering amplitude of right-handed fermion in the metric field above can be expressed as
\begin{align}\label{amp}
  i{\cal M}=i\bar{u}(P_2)\bigg[Ap_i-\frac{B}{2}i\e^{ijk}\s^j q_k\bigg]u(P_1)h_{0i}(Q),
\end{align}
with $A$ and $B$ terms to be identified as couplings of energy and spin to metric respectively. The $A$ term comes from modified dispersion relation by the metric perturbation $p_0\to p_0+\d p_0$ satisfying
\begin{align}
  (P^{\m}+\d p_0\d^{\m0})(P^{\n}+\d p_0\d^{\n0})(g_{\m\n}+h_{\m\n})=0\to\d p_0=-p_iv_i.
\end{align}
On the other hand, we note that the Fourier transform of \eqref{vorticity} gives $\tilde{\o}^j=-\frac{i}{2}\e^{ijk}q_k\tilde{v}_i$ (we use $\widetilde{}$ for quantities in momentum space), and also $S^i=\x_2^\dg\frac{\s^i}{2}\x_1$ (at $O(q)$ we can ignore the difference between $\x_2$ and $\x_1$), therefore the $B$ term give the following potential $-\vec{ S}\cdot\vec{ \o}$, which is the well-known spin-vorticity coupling. Similar conclusion can be obtained for left-handed fermion.

For massive fermions, we know $A(Q^2=0)=1$, $B(Q^2=0)=\frac{1}{2}$. These results do not renormalize by radiative corrections \cite{Kobzarev:1962wt,Pagels:1966zza}, i.e. the equivalence principle holds. On the other hand, the massless limit is known to be continuous in spin kinetic theory for massive fermions \cite{Hattori:2019ahi,Weickgenannt:2019dks,Gao:2019znl,Liu:2020flb,Guo:2020zpa}. Thus we expect non-renormalization of the GFF remains valid for massless fermions. Since the equivalence principle requires Lorentz invariance, which is lost in a medium, we expect that non-renormalization of the GFF to be violated for fermions interacting with medium. In the next section, we will study radiative correction to the GFF in medium.

\section{Medium correction to gravitational form factors}\label{sec_HTL}

In order to consider generalization of the GFF in a finite temperature medium, we face a conceptual problem: dissipative effect in medium will invalidate the scattering picture in defining the GFF. In vacuum the scattering amplitude can be expressed equivalently by LSZ reduction formula as correlation function with external momenta taking on-shell limit. Since correlation function is still well-defined, we take correlation function as generalized definition of the GFF in medium. The correction to the GFF reduces to correction to graviton-fermion vertex, i.e. amputated three point correlation function. An extra complication is that in vacuum time-ordering is sufficient, while in finite temperature medium we need to specify the operator ordering. We use the $ra$-basis in real-time formalism of finite temperature field theory. Using metric field as an example, the fields in $ra$-basis and the counterpart in Schwinger-Keldysh contour are related as $h_{\m\n,r}=\frac{1}{2}(h_{\m\n,1}+h_{\m\n,2})$, $h_{\m\n,a}=h_{\m\n,1}-h_{\m\n,2}$. Here $r$ and $a$ fields correspond to background and fluctuation fields respectively. We take the external leg operators as $\bar{\j}_r(P_1)$, $\j_a(P_2)$ and $h_{\m\n,r}(Q)$, and denote the corresponding amputated correlation function by $\d\G^{\m\n}$. Choosing $h_{\m\n,r}$ ensures that the metric field can serve as a background, and the choices of fermion operators are not unique. With the choice above, we can view $\d\G^{\m\n}h_{\m\n}$ as the fermion advanced self-energy in the background metric field.

Below we calculate $\d\G^{\m\n}$ in quantum electrodynamics (QED) plasma as an example. In the medium, $\d\G^{\m\n}$ contains both real and imaginary parts. The real part an be interpreted as potential (as we shall see, this is not always true) while the imaginary part corresponds to dissipative effect of the medium. Since the spin-vorticity coupling of our interest belongs to potential, we will retain only real part in the calculation. The content below will be divided into four parts: in subsection~\ref{subsec_struc} we list the main procedures of the calculation and display the structure of the results; in subsection~\ref{subsec_coll} we show potential collinear divergence and its cancellation; in subsection~\ref{subsec_corr} we perform the phase space integrals and convert the results into GFF; in subsection~\ref{subsec_disc}, we discuss medium correction to spin-vorticity coupling based on the results of GFF, and clarify its relation and difference with ACVE. We will also discuss the infrared divergence and its regularization in medium.

\subsection{Structure of form factors}\label{subsec_struc}

\begin{figure}[t]
\includegraphics[width=0.4\textwidth]{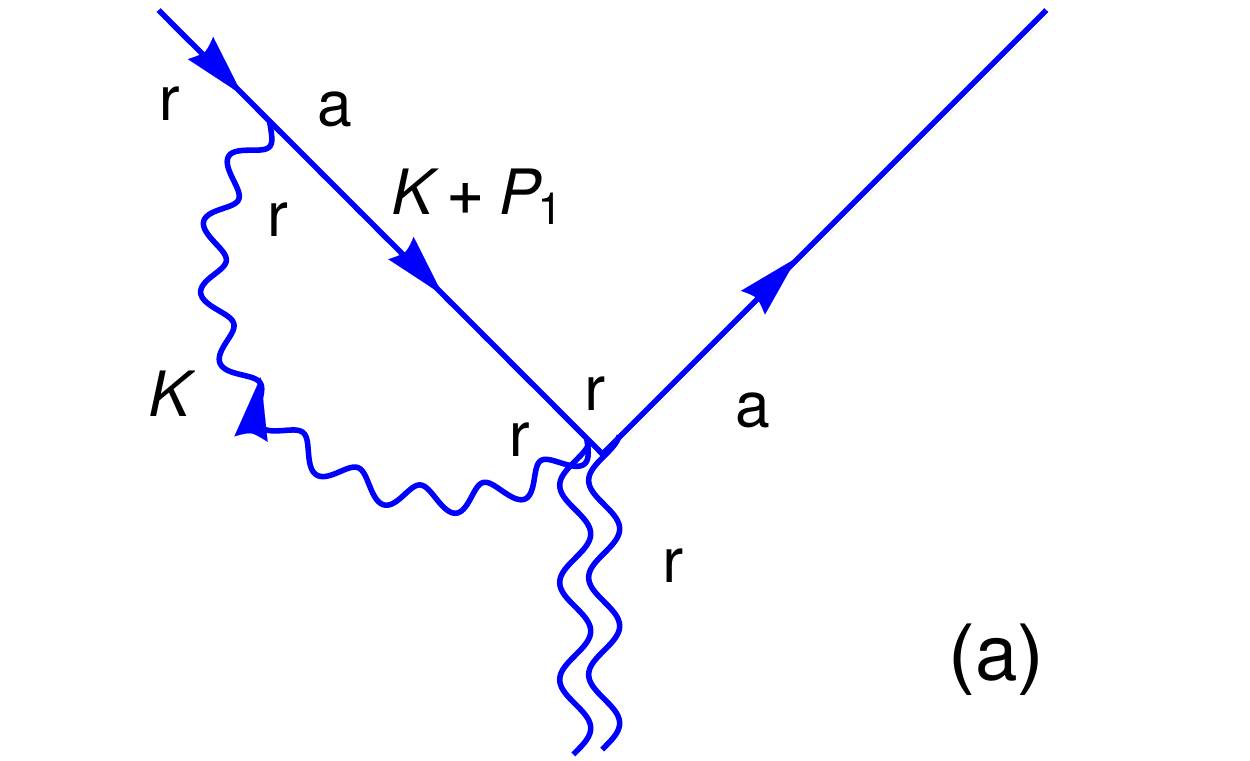}
\includegraphics[width=0.4\textwidth]{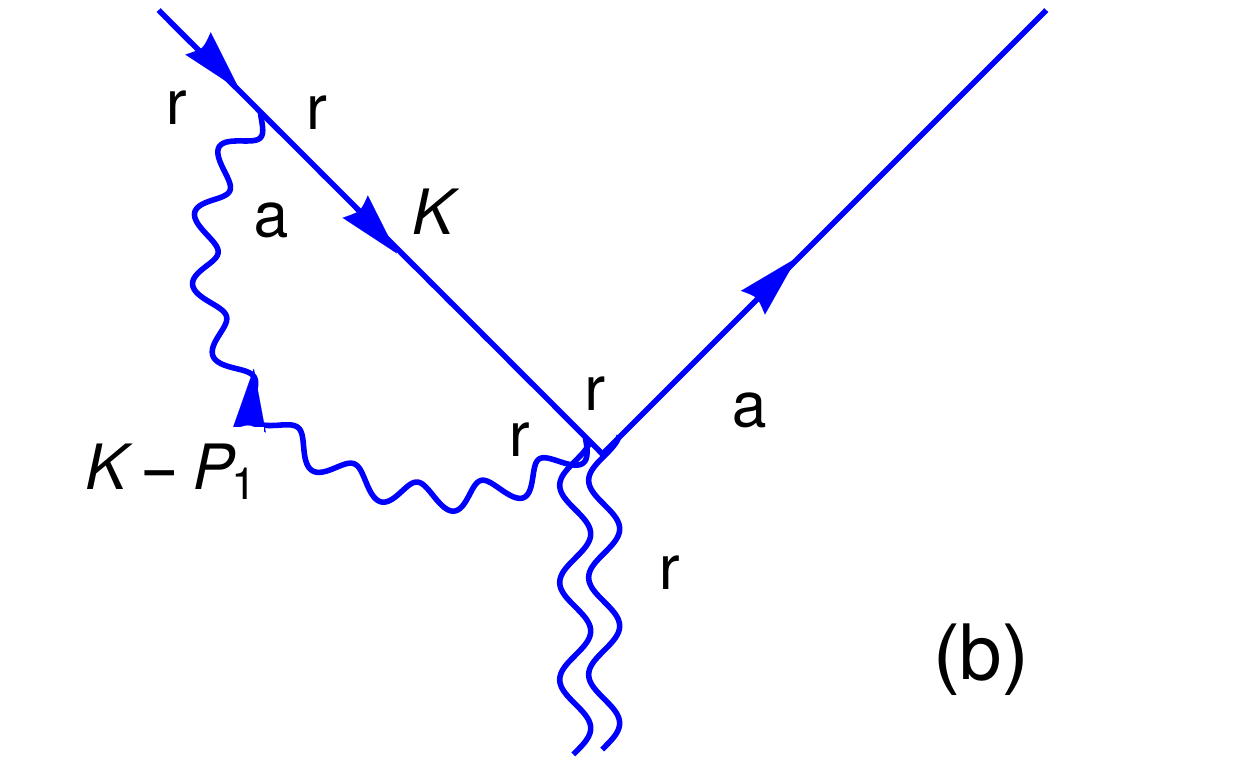}
\caption{(color online) Radiative correction diagrams containing graviton-photon-fermion vertex, with arrows indicating direction of momenta. Two similar diagrams with photon propagator connecting to the other external leg not shown.}\label{fig_diag1}
\end{figure}

\begin{figure}[t]
\includegraphics[width=0.3\textwidth]{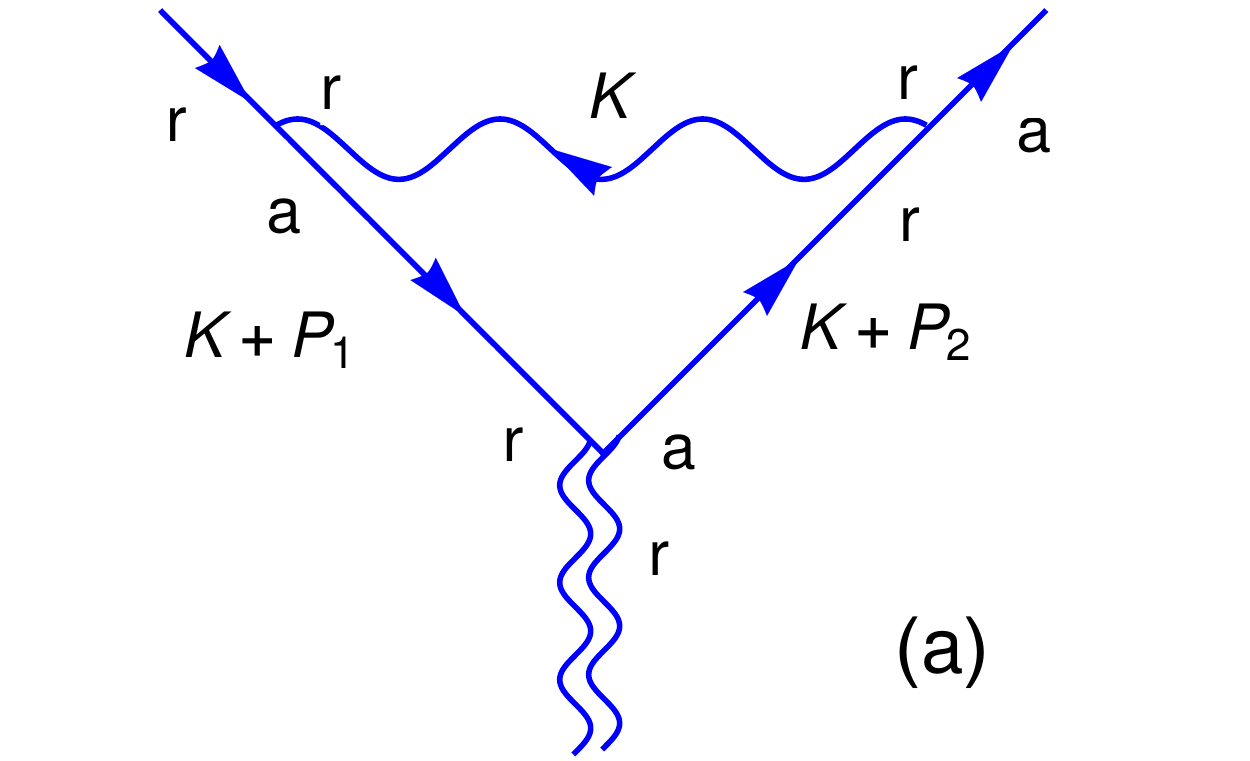}
\includegraphics[width=0.3\textwidth]{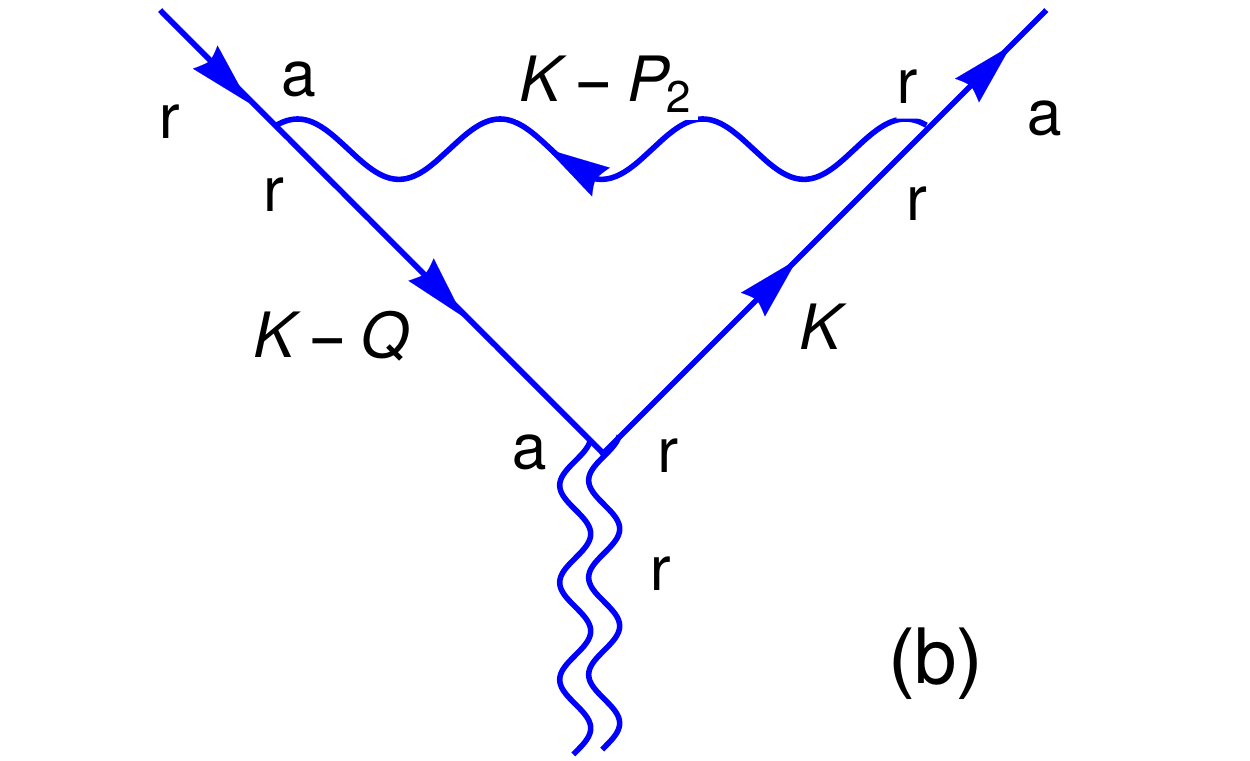}
\includegraphics[width=0.3\textwidth]{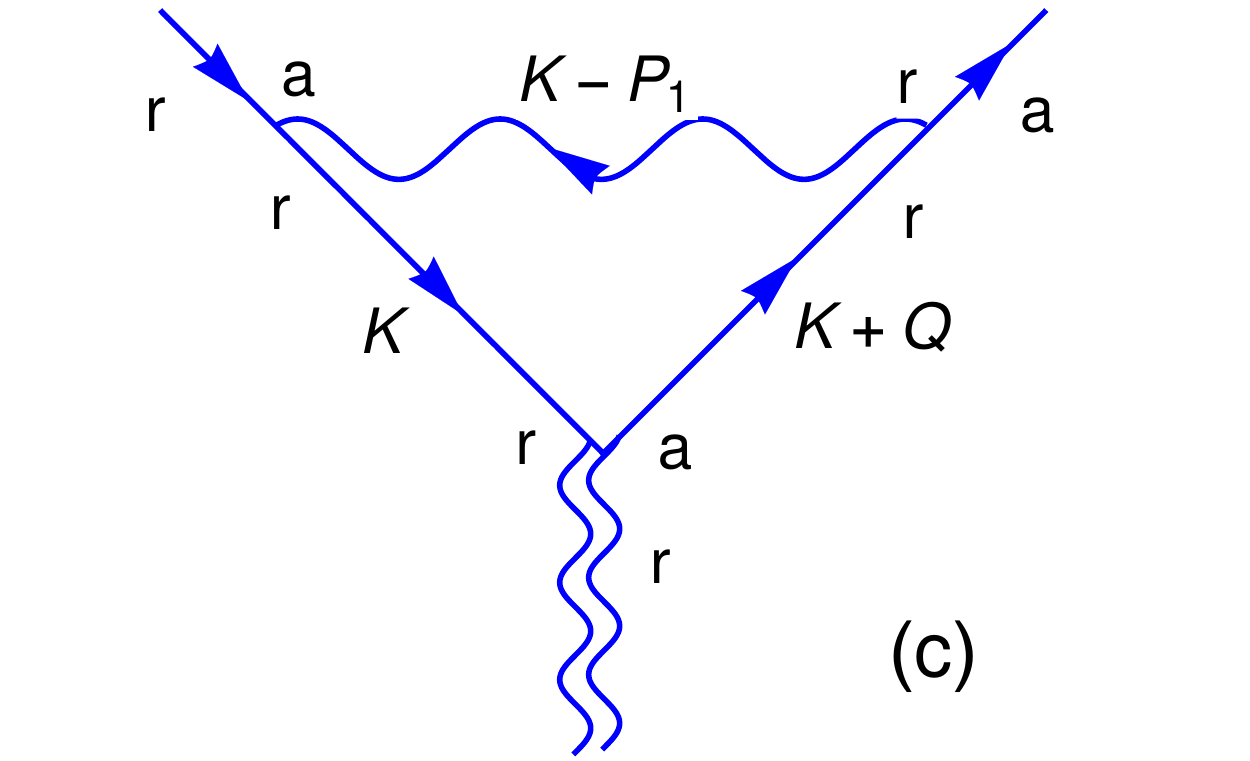}
\caption{(color online) Radiative correction diagrams containing graviton-fermion vertex, with arrows indicating direction of momenta.}\label{fig_diag2}
\end{figure}

\begin{figure}[t]
\includegraphics[width=0.3\textwidth]{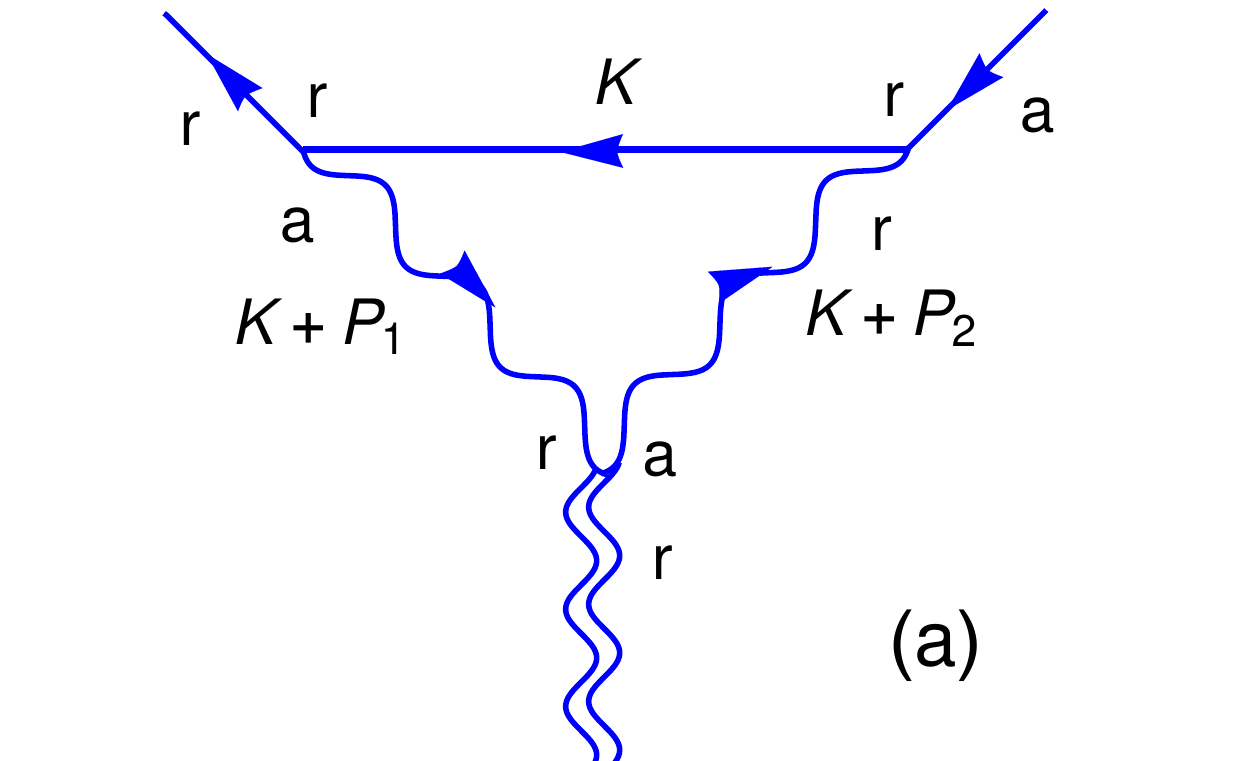}
\includegraphics[width=0.3\textwidth]{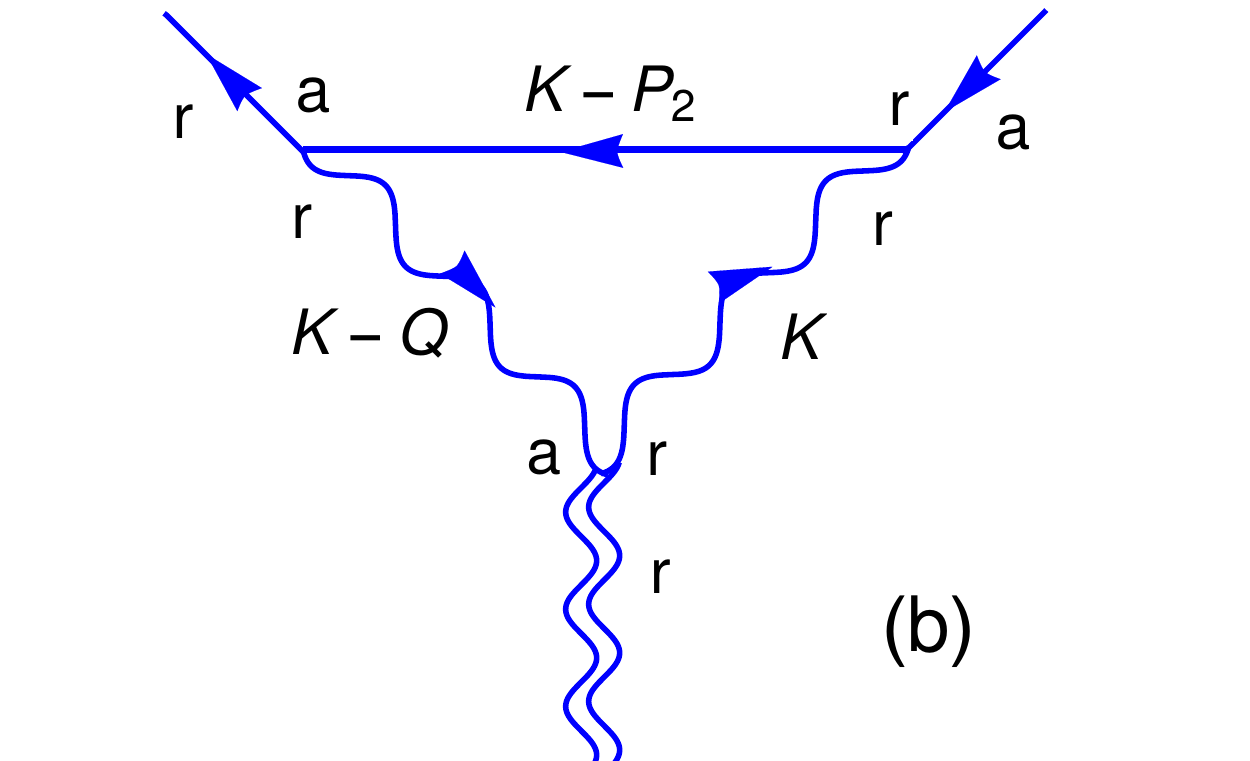}
\includegraphics[width=0.3\textwidth]{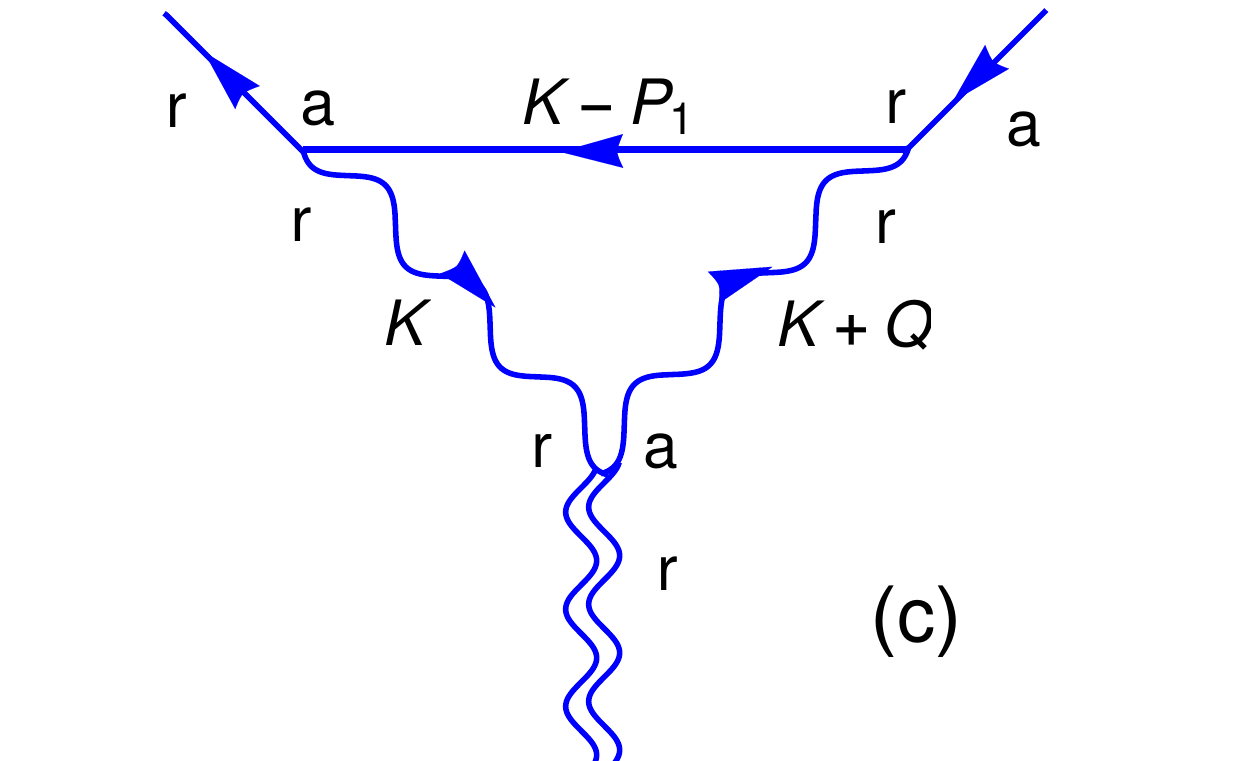}
\caption{(color online) Radiative correction diagrams containing graviton-photon vertex, with arrows indicating direction of momenta.}\label{fig_diag3}
\end{figure}

We first draw three classes of diagrams at one-loop level, with Figs.~\ref{fig_diag1}, \ref{fig_diag2} and \ref{fig_diag3} involving fermion-photon vertex, fermion-graviton vertex and photon-graviton vertex respectively. We have not shown fermion self-energy diagram, which will be treated separately. To simplify the calculations, we consider fermion momenta much less than the medium temperature: $P_{1,2}\ll T$ and use the HTL approximation, i.e. we keep only leading contributions in temperature. Although the kinematic constraint is not phenomenologically well-motivated, it allows us to display clearly the medium correction to spin-vorticity coupling.

We need the following propagators in $ra$-basis \cite{Chou:1984es}
\begin{align}\label{props}
  &S_{ra}(P)=\frac{i{\slashed P}}{P^2+i\e\,\sgn(p_0)},\quad S_{ar}(P)=\frac{i{\slashed P}}{P^2-i\e\,\sgn(p_0)},\quad S_{rr}(P)={\slashed P}2\p\e(p_0)\(\frac{1}{2}-\tilde{f}(p_0)\)\d(P^2),\nonumber\\
  &D_{ra}^{\m\n}(Q)=\frac{-ig^{\m\n}}{Q^2+i\e\,\sgn(q_0)},\quad D_{ar}^{\m\n}(Q)=\frac{-ig^{\m\n}}{Q^2-i\e\,\sgn(q_0)},\quad D_{rr}^{\m\n}(Q)=-2\p\e(q_0)\(\frac{1}{2}+f(q_0)\)g^{\m\n}.
\end{align}
The vertices involving graviton are given by
\begin{align}\label{vertices}
  &\frac{\d T^{\m\n}}{\d\bar{\j}(K_1)\d\j(K_2)}=\frac{\g^{\{\m}(K_1-K_2)^{\n\}}}{2},\nonumber\\
  &\frac{\d T^{\m\n}}{\d\bar{\j}(K_1)\d\j(K_2)\d A_\r(Q)}=-e\g^{\{\m} g^{\n\}\r},\nonumber\\
  &\frac{\d T^{\m\n}}{\d A_\r(K_1)\d A_\s(K_2)}=\big[K_1^\m K_2^\n g_{\r\s}-K_1^\m K_{2\r}\d^\n_\s-K_{1\s}K_2^\n\d^\m_\r+K_1\cdot K_2\d^\m_\r\d^\n_\s-\frac{1}{2}g^{\m\n}(K_1\cdot K_2g_{\r\s}-K_{1\s}K_{2\r})\nonumber\\
    &+(\m\leftrightarrow \n)\big].
\end{align}
Here all the field momenta flow into the vertices. We take fermion charge to be $-e$. Note that we do not specify the $ra$-labelings of the fields. \eqref{vertices} apply to any permutation containing only one $a$-field. For example, the first line of \eqref{vertices} applies to one of the following cases: $h_{\m\n}^r\bar\j^a\j^r$, $h_{\m\n}^r\bar\j^r\j^a$, $h_{\m\n}^a\bar\j^r\j^r$.

We first consider contribution from Fig.~\ref{fig_diag1}. The left diagram is given by
\begin{align}\label{diag1_left}
  &\int_K(-e\g^{\{\m}g^{\n\}\r})\frac{i({\slashed K}+{\slashed P}_1)}{(K+P_1)^2}(-ie\g^\s)2\p\d(K^2)f(k_0)\nonumber\\
  \simeq&e^2\int_K2\p\d(K^2)\(2K^{\{\m}\g^{\n\}}-{\slashed K}g^{\m\n}\)\frac{1}{2K\cdot P_1}f(k_0),
\end{align}
where we have used the HTL approximation in the numerator ${\slashed K}+{\slashed P}_1\simeq{\slashed K}$ and further used the on-shell condition $P_1^2=0$ to simplify the denominator. Note that we have dropped the $i\e$ in the denominator, which does not affect the result. The reason is $i\e$ plays a role only when the denominator approaches zero. We will see in the next subsection that collinear divergence from the vanishing of this denominator cancels out completely. The contribution from the right diagram can be worked out similarly. The only difference is a replacement of the distribution function $f(k_0)\to\tilde{f}(k_0)$\footnote{The distribution functions from the propagators give the following replacement: $f(k_0)\to-\tilde{f}(k_0)$。The extra minus sign is canceled by another one in $\frac{1}{(K-P_1)^2}=-\frac{1}{2K\cdot P_1}$.} Apart from the two diagrams in Fig.~\ref{fig_diag1}, the photon propagators can also been attached to external lines with momentum $P_2$. The corresponding contribution can be obtained from the above by the replacement $P_1\to P_2$. Collecting the above contributions, we have
\begin{align}\label{diag1}
  e^2\int_K\(2K^{\{\m}\g^{\n\}}-{\slashed K}g^{\m\n}\)2\p\d(K^2)\(\frac{1}{2K\cdot P_1}+\frac{1}{2K\cdot P_2}\)\(f(k_0)+\tilde{f}(k_0)\).
\end{align}
Note that the factor $\d(K^2)$ indicates the integral above comes from two contributions at $k_0=\pm k$. We can easily show by a change of variable $\vec{k}\to-\vec{k}$ that the two contributions are identical. So in the following we only need to consider twice the contribution from $k_0=k$. We can see from a simple power counting that HTL gives a contribution as $\int_K\d(K^2)\sim O(K^2)$.

Next we consider contribution from Fig.~\ref{fig_diag2}. The left diagram corresponds to the case with photon being on-shell, while the middle and right diagrams correspond to the cases with one of the fermions being on-shell. To simplify the calculations, we choose to have the on-shell particles carry momenta $K$. A simple power counting shows the leading order (LO) contribution reads $\int_K\d(K^2)K\sim O(K^3)$. Since the LO is an odd function of $K$, by the same change of variables $\vec{k}\to-\vec{k}$, it is not difficult to find that the LO contributions from $k_0=\pm k$ cancel entirely. Thus we have to consider next-to-leading order (NLO) contribution. We keep the NLO contributions from the three diagrams respectively. The contribution from the left diagram reads
\begin{align}\label{diag2_left}
  &\int_K(-ie\g^\r)\frac{i({\slashed K}+{\slashed P}_2)}{(K+P_2)^2}\g^{\{\m}(K+P)^{\n\}}\frac{i({\slashed K}+{\slashed P}_1)}{(K+P_1)^2}(-ie\g^\s)(-g_{\r\s})2\p\d(K^2)f(k_0)\nonumber\\
  \simeq &e^2\int_K2\p\d(K^2)\big[2{\slashed P}_1\g^{\{\m}{\slashed K}K^{\n\}}+2{\slashed K}\g^{\{\m}{\slashed P}_2K^{\n\}}+2{\slashed K}\g^{\{\m}{\slashed K}P^{\n\}}\big]\frac{1}{2K\cdot P_1}\frac{1}{2K\cdot P_2}f(k_0).
\end{align}
Using the following identity
\begin{align}\label{gamma_identity}
  \g^\m\g^\n\g^\r=\g^\m g^{\n\r}-\g^\n g^{\m\r}+\g^\r g^{\m\n}-i\e^{\m\n\r\s}\g^5\g_\s,
\end{align}
we can simplify \eqref{diag2_left} as
\begin{align}\label{diag2_left2}
  e^2\int_K2\p\d(K^2)\big[8P^{\{\m}K^{\n\}}{\slashed K}-4K\cdot P\g^{\{\m}K^{\n\}}-2i\e^{\a\b\l\{\m}\g^5\g_\l K_\a Q_\b K^{\n\}}\big]\frac{1}{2K\cdot P_1}\frac{1}{2K\cdot P_2}f(k_0).
\end{align}
The contribution from the middle diagram is given by
\begin{align}\label{diag2_middle}
  &\int_K(-ie\g^\r){\slashed K}\g^{\{\m}\(K-\frac{Q}{2}\)^{\n\}}\frac{i({\slashed K}-{\slashed Q})}{(K-Q)^2}(-ie\g^\s)\frac{-ig_{\r\s}}{(K-P_2)^2}2\p\d(K^2)(-\tilde{f}(k_0))\nonumber\\
  \simeq&e^2\int_K2\p\d(K^2)\tilde{f}(k_0)\{\big[4Q^{\{\m}K^{\n\}}{\slashed K}-2\g^{\{\m}K^{\n\}}K\cdot Q+2i\e^{\a\b\l\{\m}\g^5\g_\l K_\a Q_\b K^{\n\}}\big]\frac{1}{2K\cdot P_2}\frac{1}{2K\cdot Q}\nonumber\\
  &-4K^\m K^\n{\slashed K}\frac{1}{2K\cdot P_2}\frac{Q^2}{(2K\cdot Q)^2}\},
\end{align}
where the two terms in the above come from expanding the denominator $-2K\cdot Q+Q^2$ to LO and NLO terms. The contribution from the right diagram can be obtained similarly as
\begin{align}\label{diag2_right}
    &\int_K(-ie\g^\r)\frac{i({\slashed K}+{\slashed Q})}{(K+Q)^2}\g^{\{\m}\(K+\frac{Q}{2}\)^{\n\}}{\slashed K}(-ie\g^\s)\frac{-ig_{\r\s}}{(K-P_1)^2}2\p\d(K^2)(-\tilde{f}(k_0))\nonumber\\
  \simeq &e^2\int_K2\p\d(K^2)\tilde{f}(k_0)\{\big[4Q^{\{\m}K^{\n\}}{\slashed K}-2\g^{\{\m}K^{\n\}}K\cdot Q-2i\e^{\a\b\l\{\m}\g^5\g_\l K_\a Q_\b K^{\n\}}\big]\frac{1}{2K\cdot P_1}\frac{1}{2K\cdot Q}\nonumber\\
  &-4K^\m K^\n{\slashed K}\frac{1}{2K\cdot P_1}\frac{Q^2}{(2K\cdot Q)^2}\}.
\end{align}
The sum of \eqref{diag2_left2}, \eqref{diag2_middle} and \eqref{diag2_right} gives
\begin{align}\label{diag2_sum0}
  &e^2\int_K2\p\d(K^2)\{\big[8P^{\{\m}K^{\n\}}{\slashed K}-4K\cdot P\g^{\{\m}K^{\n\}}-2i\e^{\a\b\l\{\m}\g^5\g_\l K_\a Q_\b K^{\n\}}\big]\frac{1}{2K\cdot P_1}\frac{1}{2K\cdot P_2}f(k_0)\nonumber\\
    &+\big[4Q^{\{\m}K^{\n\}}{\slashed K}-2\g^{\{\m}K^{\n\}}K\cdot Q\big]\(\frac{1}{2K\cdot P_1}+\frac{1}{2K\cdot P_2}\)\frac{1}{2K\cdot Q}\tilde{f}(k_0)\nonumber\\
    &-2i\e^{\a\b\l\{\m}\g^5\g_\l K_\a Q_\b K^{\n\}}\(\frac{1}{2K\cdot P_2}\frac{1}{2K\cdot P_2}\)\tilde{f}(k_0)\nonumber\\
    &-4K^\m K^\n{\slashed K}\(\frac{1}{2K\cdot P_1}+\frac{1}{2K\cdot P_2}\)\frac{Q^2}{(2K\cdot Q)^2}\tilde{f}(k_0)\}.
\end{align}
The Dirac structures appearing above are not completely independent. Let us derive a relation among them. From the equation of motion (EOM), we know the following factors appearing in the form factors vanishes identically $\({\slashed K}\g^\m{\slashed P}_1+{\slashed P}_2\g^\m{\slashed K}\)=0$. Therefore we can write down the following identity
\begin{align}\label{aid}
  \frac{1}{2}\big[\({\slashed P_1}\g^\m{\slashed K}+{\slashed K}\g^\m{\slashed P_2}\)+\({\slashed K}\g^\m{\slashed P}_1+{\slashed P}_2\g^\m{\slashed K}\)\big]=\frac{1}{2}\big[\({\slashed P_1}\g^\m{\slashed K}+{\slashed K}\g^\m{\slashed P_2}\)-\({\slashed K}\g^\m{\slashed P}_1+{\slashed P}_2\g^\m{\slashed K}\)\big].
\end{align}
Using \eqref{gamma_identity} we obtain
\begin{align}\label{aid2}
  -2K\cdot P\g^\m+2{\slashed K}P^\m=i\e^{\m\a\b\l}\g^5\g_\l K_\a Q_\b.
\end{align}
Multiplying both sides by $K^\n$ and symmetrizing the indices, we obtain
\begin{align}\label{aid3}
  -2K\cdot P\g^{\{\m}K^{\n\}}+2{\slashed K}P^{\{\m}K^{\n\}}=iK^{\{\n}\e^{\m\}\a\b\l}\g^5\g_\l K_\a Q_\b .
\end{align}
Using \eqref{aid3}, we can further simplify \eqref{diag2_sum0} as
\begin{align}\label{diag2_sum}
  &e^2\int_K2\p\d(K^2)\{\big[-2\g^{\{\m}K^{\n\}}\(\frac{1}{2K\cdot P_1}+\frac{1}{2K\cdot P_2}\)+4P^{\{\m}K^{\n\}}{\slashed K}\frac{1}{2K\cdot P_1}\frac{1}{2K\cdot P_2}\big](f(k_0)+\tilde{f}(k_0))\nonumber\\
  &+\big[4Q^{\{\m}K^{\n\}}{\slashed K}\(\frac{1}{2K\cdot P_1}+\frac{1}{2K\cdot P_2}\)\frac{1}{2K\cdot Q}+8P^{\{\m}K^{\n\}}{\slashed K}\frac{1}{2K\cdot P_1}\frac{1}{2K\cdot P_2}\nonumber\\
  &-4K^\m K^\n{\slashed K}\(\frac{1}{2K\cdot P_1}+\frac{1}{2K\cdot P_2}\)\frac{Q^2}{(2K\cdot Q)^2}\big]\tilde{f}(k_0)\}.
\end{align}

Finally we look at the contribution from Fig.~\ref{fig_diag3}. An analysis similar to Fig.~\ref{fig_diag2} shows that the contributions from $k_0=\pm k$ again cancel out entirely, so we need to keep the NLO contribution. The contribution from the left diagram reads
\begin{align}\label{diag3_left}
  &\int_K(-ie\g_\b)(-{\slashed K})(-ie\g_\a)\frac{-ig^{\a\r}}{(K+P_2)^2}\frac{-ig^{\b\s}}{(K+P_1)^2}V_{\r\s}^{\m\n}(k_1\to K+P_1,k_2\to-(K+P_2))2\p\d(K^2)(-\tilde{f}(k_0))\nonumber\\
  \simeq&e^2\int_K2\p\d(K^2)\big[8K^{\{\m}P^{\n\}}{\slashed K}-4iK^{\{\m}\e^{\n\}\a\b\l}K_\a Q_\b -8P\cdot K K^{\{\m}\g^{\n\}}\big]\frac{1}{2K\cdot P_1}\frac{1}{2K\cdot P_2}\tilde{f}(k_0)\nonumber\\
  =&0.
\end{align}
Here we have used $V_{\r\s}^{\m\n}(k_1,k_2)$ to denote the photon-graviton vertex in \eqref{vertices}. We have used \eqref{aid3} in the last equality. The contribution from the middle diagram reads
\begin{align}\label{diag3_middle}
  &\int_K(-ie\g_\b)\frac{-i({\slashed K}-{\slashed P}_2)}{(K-P_2)^2}(-ie\g_\a)\frac{-ig^{\a\r}}{(K-Q)^2}(-g^{\b\s})V_{\r\s}^{\m\n}(k_1\to K-Q,k_2\to-K)2\p\d(K^2)f(k_0)\nonumber\\
  \simeq&e^2\int_K2\p\d(K^2)f(k_0)\{-4\big[-K^{\{\m}P_2^{\n\}}{\slashed K}-K^{\{\n}\g^{\m\}}K\cdot P_2+\frac{1}{2}g^{\m\n}K\cdot P_2{\slashed K}\big]\frac{1}{2K\cdot P_2}\frac{1}{2K\cdot Q}\nonumber\\
  -&4K^\m K^\n{\slashed K}\frac{1}{2K\cdot P_2}\frac{Q^2}{(2K\cdot Q)^2}
  -2\big[-Q^{\{\m}K^{\n\}}{\slashed K}+i K^{\{\n}\e^{\m\}\a\b\l}\g^5\g_\l Q_\a K_\b+K^{\{\n}\g^{\m\}}K\cdot Q\big]\frac{1}{2K\cdot P_2}\frac{1}{2K\cdot Q}\}\nonumber\\
  =&e^2\int_K2\p\d(K^2)\{\big[8K^{\{\m}P^{\n\}}{\slashed K}+4K^{\{\m}Q^{\n\}}{\slashed K}-2g^{\m\n}K\cdot P_2{\slashed K}\big]\frac{1}{2K\cdot P_2}\frac{1}{2K\cdot Q}f(k_0)\nonumber\\
  -&4K^\m K^\n{\slashed K}\frac{1}{2K\cdot P_2}\frac{Q^2}{(2K\cdot Q)^2}f(k_0)\}.
\end{align}
And the contribution from the right diagram is obtained similarly as
\begin{align}\label{diag3_right}
  &\int_K(-ie\g_\b)\frac{-i({\slashed K}-{\slashed P}_1)}{(K-P_1)^2}(-ie\g_\a)(-g^{\a\r})\frac{-ig^{\b\s}}{(K+Q)^2}V_{\r\s}^{\m\n}(k_1\to K,k_2\to-(K+Q))2\p\d(K^2)f(k_0)\nonumber\\
  \simeq&e^2\int_K2\p\d(K^2)\{\big[-8K^{\{\m}P^{\n\}}{\slashed K}+4K^{\{\m}Q^{\n\}}{\slashed K}+2g^{\m\n}K\cdot P_1{\slashed K}\big]\frac{1}{2K\cdot P_1}\frac{1}{2K\cdot Q}f(k_0)\nonumber\\
  -&4K^\m K^\n{\slashed K}\frac{1}{2K\cdot P_1}\frac{Q^2}{(2K\cdot Q)^2}f(k_0)\}.
\end{align}
The sum of \eqref{diag3_left}, \eqref{diag3_middle} and \eqref{diag3_right} gives
\begin{align}\label{diag3_sum}
  &e^2\int_K2\p\d(K^2)\{-8K^{\{\m}P^{\n\}}{\slashed K}\frac{1}{2K\cdot P_1}\frac{1}{2K\cdot P_2}+4K^{\{\m}Q^{\n\}}{\slashed K}\(\frac{1}{2K\cdot P_1}+\frac{1}{2K\cdot P_2}\)\frac{1}{2K\cdot Q}\nonumber\\
  -&4K^\m K^\n{\slashed K}\(\frac{1}{2K\cdot P_1}+\frac{1}{2K\cdot P_2}\)\frac{Q^2}{(2K\cdot Q)^2}\}f(k_0).
\end{align}

\subsection{Cancellation of collinear divergence}\label{subsec_coll}

Because $K$, $P_1$ and $P_2$ are light-like momenta, the factors $\frac{1}{2K\cdot P_1}$ and $\frac{1}{2K\cdot P_2}$ in the integrand obtained in the previous subsection both lead to divergence. Using the former as an example, we take the angle between $\vec{k}$ and $\vec{p}_1$ to be $\th$, and the integration of $\th$ can be written as
\begin{align}
  \int_{-1}^1 d\cos\th\frac{1}{2K\cdot P_1}=\int_{-1}^1 d\cos\th\frac{1}{2kp_1(1-\cos\th)}.
\end{align}
When $\cos\th\to1$, the above integral contains logarithmic divergence. Since this divergence occurs when $K$ and $P_1$ become parallel, we refer to it as collinear divergence. Keeping the $i\e$ ignored in the above would turn the result into $\ln\e$, which still fails to treat the divergence. If we consider thermal masses of photon and fermion in the medium, the divergence can be effectively cut off by the thermal masses. In fact, here we do not have to introduce thermal masses. We shall show, all the collinear divergences cancel out entirely in the final GFF.

First of all, we note that the following Dirac structures appear in the previous subsection: $K^{\{\m}\g^{\n\}}$, ${\slashed K}g^{\m\n}$, ${\slashed K}K^{\{\m}P^{\n\}}$, ${\slashed K}K^{\{\m}P^{\n\}}$ and ${\slashed K}K^{\m}K^{\n}$. Among them the first structure cancels out in the sum of \eqref{diag1} and \eqref{diag2_sum}, and all the remaining structures contain ${\slashed K}$. Since the collinear divergence occurs when $K$ and $P_{1,2}$ become parallel, in this case ${\slashed K}\propto{\slashed P}_{1,2}$. We can see from EOM that they do not contribute to the GFF.

Next we classify the remaining terms according to the Dirac structures as follows
\begin{align}\label{finite}
  I:\;\int_K2\p\d(K^2)(-{\slashed K}g^{\m\n})\(\frac{1}{2K\cdot P_1}+\frac{1}{2K\cdot P_2}\)\(f(k_0)+\tilde{f}(k_0)\),\nonumber\\
  II:\;\int_K2\p\d(K^2)4P^{\{\m}K^{\n\}}{\slashed K}\(\frac{1}{2K\cdot P_1}-\frac{1}{2K\cdot P_2}\)\frac{1}{2K\cdot Q}\(f(k_0)+\tilde{f}(k_0)\),\nonumber\\
  III:\;\int_K2\p\d(K^2)4Q^{\{\m}K^{\n\}}{\slashed K}\(\frac{1}{2K\cdot P_1}+\frac{1}{2K\cdot P_2}\)\frac{1}{2K\cdot Q}\(f(k_0)+\tilde{f}(k_0)\),\nonumber\\
  IV:\;\int_K2\p\d(K^2)\(-4K^{\m}K^{\n}\){\slashed K}\(\frac{1}{2K\cdot P_1}+\frac{1}{2K\cdot P_2}\)\frac{Q^2}{(2K\cdot Q)^2}\(f(k_0)+\tilde{f}(k_0)\).
\end{align}
Interestingly the results above contain a common factor $f(k_0)+\tilde{f}(k_0)$. The same factor contributes to the thermal mass of fermion.

\subsection{Medium correction to form factors}\label{subsec_corr}

In this subsection, we will perform the phase space integrations to obtain explicit expressions for the GFF. Since we are interested in the coupling of spin and static vorticity, we further require $q_0=0$, i.e. there is no energy exchange but only momentum exchange between fermion and graviton in the medium frame. In this case, we can parametrize the 3-momenta in the medium frame as
\begin{align}\label{para}
  &\vec{p}_1=\(0,-\frac{q}{2},p\),\nonumber\\
  &\vec{p}_2=\(0,\frac{q}{2},p\),\nonumber\\
  &\vec{q}=\(0,q,0\),\nonumber\\
  &\vec{k}=k\(\sin\th\cos\vj,\sin\th\sin\vj,\cos\th\).
\end{align}
Note that when $q_0=0$, we have $P\cdot Q=-\vec{p}\cdot\vec{q}=0$. Without loss of generality, we choose to have $p$ and $q$ pointing along $z$ and $y$ respectively in the above. Below we calculate $I$-$IV$ in \eqref{finite}. We first simplify the integrand of $I$
\begin{align}\label{I}
  I:\;&(-{\slashed K}g^{\m\n})\(\frac{1}{2K\cdot P_1}+\frac{1}{2K\cdot P_2}\)\(f(k_0)+\tilde{f}(k_0)\)\nonumber\\
  =&-\g_\l g^{\m\n}\(K^\l-\frac{K\cdot u}{P_1\cdot u}P_1^\l\)\frac{1}{2K\cdot P_1}(f(k_0)+\tilde{f}(k_0))+(P_1\to P_2).
\end{align}
Here $u^\m=(1,0,0,0)$ is the medium frame vector. In the second line, we have subtracted a contribution proportional to ${\slashed P}_1$, eliminating the collinear divergence, which does not contribute to the GFF. From $u_\l\(K^\l-\frac{K\cdot u}{P_1\cdot u}P_1^\l\)=0$, we can see $\l$ can only be spatial indices. From our parametrization and rotational invariance, $\l$ can only be $z$ and $y$. Moreover by the EOM ${\slashed P}_{1,2}=0$. The difference of them gives ${\slashed Q}=0$. Since $Q=(0,0,q,0)$, we conclude that $\l=y$ also does not contribute to the GFF. When $\l=z$, the integration of \eqref{I} can be calculated as
\begin{align}\label{I_z}
  \;&\int_K2\p\d(K^2)\(K^\l-\frac{K\cdot u}{P_1\cdot u}P_1^\l\)\frac{1}{2K\cdot P_1}\(f(k_0)+\tilde{f}(k_0)\)\nonumber\\
  =&a\int d\cos\th d\vj\(\cos\th-\frac{p}{(p^2+\frac{q^2}{4})^{1/2}}\)\(\frac{1}{2\((p^2+\frac{q^2}{4})^{1/2}-p\cos\th\)}\)\nonumber\\
  =&-4\p a\frac{1}{2p},
\end{align}
with $a=e^2\int\frac{kdk}{(2\p)^3}\(f(k_0)+\tilde{f}(k_0)\)$. We have kept only the dominant terms in the limit $q\to0$. Similarly we can obtain an identical result from $(P_1\to P_2)$. Collecting the two parts and rewriting the result into a covariant form, we have
\begin{align}\label{I_exp}
  I=4\p a\g\cdot\hat{p}g^{\m\n}\frac{1}{p}.
\end{align}
Next we consider $II$. Similar to the treatment of $I$, we first rewrite the integrand of $II$ as
\begin{align}\label{II}
  &4P^{\{\m}K^{\n\}}{\slashed K}\(\frac{1}{2K\cdot P_1}-\frac{1}{2K\cdot P_2}\)\frac{1}{2K\cdot Q}\(f(k_0)+\tilde{f}(k_0)\)\nonumber\\
  =&4\g_\l\big[\(K^\l-\frac{K\cdot u}{P_1\cdot u}P_1^\l\)\(P^{\{\m}K^{\n\}}-\frac{K\cdot u}{P_1\cdot u}P^{\{\m}P_1^{\n\}}\)+\(K^\l-\frac{K\cdot u}{P_1\cdot u}P_1^\l\)\frac{K\cdot u}{P_1\cdot u}P^{\{\m}P_1^{\n\}}\big]\nonumber\\
  \times&\frac{1}{2K\cdot P_1}\frac{1}{2K\cdot Q}\(f(k_0)+\tilde{f}(k_0)\)
-(P_1\to P_2).
\end{align}
The first and second terms in the square bracket above are tensor and vector of $K$ respectively. The indices of the first term can only take spatial one. By rotational invariance, $\l\n$ can be decomposed into the following tensor structures $\d_{ij}$, $\hat{p}_i\hat{p}_j$, $\hat{q}_i\hat{q}_j$ and $(\hat{p}_i\hat{q}_j+\hat{q}_i\hat{p}_j)$. For reasons discussed earlier, $\hat{q}_i\hat{q}_j$ does not contribute to the GFF. To ease the calculations below, we define $\hat{l}_i=\e^{ijk}\hat{q}_j\hat{p}_k$ and use $\hat{l}_i\hat{l}_j$ to replace $\d_{ij}$ in the tensor basis above. This leads to the simple correspondences: $\hat{l}_i\hat{l}_j=xx$, $\hat{p}_i\hat{p}_j=zz$, $(\hat{p}_i\hat{q}_j+\hat{q}_i\hat{p}_j)=yz$. Below we show the integration appearing in the first term of \eqref{II}\footnote{Note that the integrand contains $\frac{1}{2K\cdot Q}=\frac{1}{-2kq\sin\th\sin\vj}$. Superficially this factor has a non-integrable divergence as $\vj\to0$, but this divergence cancels out when we sum over contributions from $\vj$ and $\vj+\p$.}
\begin{align}\label{III_integrals}
  &\int_K2\p\d(K^2)\(K^\l-\frac{K\cdot u}{P_1\cdot u}P_1^\l\)\(K^{\n}-\frac{K\cdot u}{P_1\cdot u}P_1^{\n}\)\frac{1}{2K\cdot P_1}\frac{1}{2K\cdot Q}\(f(k_0)+\tilde{f}(k_0)\)\nonumber\\
  &=\left\{\begin{array}{l@{\quad\quad}l}
  4\p a\frac{\ln\frac{2p}{q}}{8p^2},& \l\n=xx\\
  4\p a\frac{1}{8p^2},& \l\n=zz\\
  4\p a\frac{1}{pq},&   \l\n=zy
\end{array}
\right.
\end{align}
Again we have kept only the dominant terms as $q\to0$. The calculation of the second term is similar to that of $I$. We can fix $\l=z$. Keeping the dominant terms in the limit $q\to0$ means $\n=0$ or $z$. The two cases give the following identical result
\begin{align}
  &\int_K2\p\d(K^2)\(K^\l-\frac{K\cdot u}{P_1\cdot u}P_1^\l\)\(\frac{K\cdot u}{P_1\cdot u}P_1^{\n}\)\frac{1}{2K\cdot P_1}\frac{1}{2K\cdot Q}\(f(k_0)+\tilde{f}(k_0)\)\nonumber\\
  =&4\p a\frac{-\ln\frac{2p}{q}}{8p^2}.
\end{align}
The other contribution $-(P_1\to P_2)$ gives an identical contribution for $\l\n=xx$ and $\l\n=zz$, but gives an opposite one for $\l\n=yz$. Collecting the above, we obtain
\begin{align}\label{II_exp}
  II=4\p a\big[\g\cdot\hat{l}P^{\{\m}\hat{l}^{\n\}}\frac{\ln\frac{2p}{q}}{p^2}+\g\cdot\hat{p}P^{\{\m}\hat{p}^{\n\}}\frac{1}{p^2}-\g\cdot\hat{p}P^\m P^\n\frac{\ln\frac{2p}{q}}{p^3}\big].
\end{align}
$III$ does not require extra calculations. We use the existing results in \eqref{III_integrals} and note that $(P_1\to P_2)$ gives an identical contribution now to arrive at
\begin{align}\label{III_exp}
  III=4\p a\big[2\(\g\cdot\hat{p}\hat{q}^\m\hat{q}^\n+\g\cdot\hat{q}\hat{q}^{\{\m}\hat{p}^{\n\}}\)\frac{1}{p}\big].
\end{align}
The calculations of $IV$ are more involved. We follow the treatment above to rewrite the integrand as
\begin{align}\label{IV}
  &-4K^\m K^\n{\slashed K}\(\frac{1}{2K\cdot P_2}+\frac{1}{2K\cdot P_1}\)\frac{Q^2}{(2K\cdot Q)^2}\(f(k_0)+\tilde{f}(k_0)\)\nonumber\\
  =&-4\g_\l\(K^\l-\frac{K\cdot u}{P_1\cdot u}P_1^\l\)\(K^\m-\frac{K\cdot u}{P_1\cdot u}P_1^\m\)\(K^\n-\frac{K\cdot u}{P_1\cdot u}P_1^\n\)\frac{1}{2K\cdot P_1}\frac{Q^2}{(2K\cdot Q)^2}\(f(k_0)+\tilde{f}(k_0)\)\nonumber\\
  &-4\g_\l\(K^\l-\frac{K\cdot u}{P_1\cdot u}P_1^\l\)\big[\(K^\m-\frac{K\cdot u}{P_1\cdot u}P_1^\m\)\frac{K\cdot u}{P_1\cdot u}P_1^\n+(\m\to\n)\big]\frac{1}{2K\cdot P_1}\frac{Q^2}{(2K\cdot Q)^2}\(f(k_0)+\tilde{f}(k_0)\)\nonumber\\
  &-4\g_\l\(K^\l-\frac{K\cdot u}{P_1\cdot u}P_1^\l\)\frac{K\cdot u}{P_1\cdot u}P_1^\m\frac{K\cdot u}{P_1\cdot u}P_1^\n\frac{1}{2K\cdot P_1}\frac{Q^2}{(2K\cdot Q)^2}\(f(k_0)+\tilde{f}(k_0)\)+(P_1\to P_2).
\end{align}
Similar to the earlier analysis, the rank-three tensor indices can take $\l\m\n=zzz,\;zxx,\;yxx,\;zyy,\;zzy$, rank-two tensor indices can take $\l\m=zz,\;xx,\;zy$ and vector index takes $\l=z$. The remaining indices can only take $0$ or $z$ in the limit $q\to0$. Each term in the above contains $\frac{1}{(2K\cdot Q)^2}=\frac{1}{4k^2q^2\sin^2\th\sin^2\vj}$. This time we find the divergence from integration of $\vj$ is unavoidable, so we have to keep $i\e$ to regularize the divergence. The regularized results contain both real and imaginary parts, with the imaginary part describing dissipative effect. Since the spin-vorticity coupling potential of our interest comes from the real part, we will discuss the real part of the results only. The calculational details are elaborated in the appendix. Here we list only the main results. The first term involves the following integral.
\begin{align}\label{IV_integral}
  &\int_K2\p\d(K^2)\(K^\l-\frac{K\cdot u}{P_1\cdot u}P_1^\l\)\(K^\m-\frac{K\cdot u}{P_1\cdot u}P_1^\m\)\(K^\n-\frac{K\cdot u}{P_1\cdot u}P_1^\n\)\frac{1}{2K\cdot P_1}\frac{Q^2}{(2K\cdot Q)^2}\(f(k_0)+\tilde{f}(k_0)\)\nonumber\\
  &=\left\{\begin{array}{l@{\quad\quad}l}
  -4\p a\frac{1}{4p},& \l\m\n=zzz\\
  -4\p a\frac{1}{8p},& \l\m\n=zxx\\
  4\p a\frac{1}{8p}& \l\m\n=zyy
\end{array}
\right.
\end{align}
The remaining structures are suppressed in the limit $q\to0$. The second term involves the following integrals (keeping only the unsuppressed choices of indices)
\begin{align}\label{IV_integral2}
  &\int_K2\p\d(K^2)\(K^\l-\frac{K\cdot u}{P_1\cdot u}P_1^\l\)\big[\(K^\m-\frac{K\cdot u}{P_1\cdot u}P_1^\m\)\frac{K\cdot u}{P_1\cdot u}P_1^\n+(\m\to\n)\big]\frac{1}{2K\cdot P_1}\frac{Q^2}{(2K\cdot Q)^2}\(f(k_0)+\tilde{f}(k_0)\)\nonumber\\
  &=\left\{\begin{array}{l@{\quad\quad}l}
  4\p a\frac{1}{4p},& \l\m\n=zzz\\
  4\p a\frac{\ln\frac{2p}{q}}{8p},& \l\m\n=xxz
\end{array}
\right.
\end{align}
The results with $\n=0$ are identical to the above. The third term involves the following integrals
\begin{align}\label{IV_integral3}
  &\int_K2\p\d(K^2)\(K^\l-\frac{K\cdot u}{P_1\cdot u}P_1^\l\)\frac{K\cdot u}{P_1\cdot u}P_1^\m\frac{K\cdot u}{P_1\cdot u}P_1^\n\frac{1}{2K\cdot P_1}\frac{Q^2}{(2K\cdot Q)^2}\(f(k_0)+\tilde{f}(k_0)\)\nonumber\\
  =&-4\p a\frac{1}{8p},\quad \l\m\n=zzz.
\end{align}
The results with $\m\n=0z$ and $00$ are identical to the above. We can then rewrite the result of $IV$ into the following covariant form
\begin{align}\label{IV_exp}
  IV=&4\p a\big[\(\g\cdot\hat{p}\hat{l}^\m\hat{l}^\n+2\g\cdot\hat{l}\hat{l}^{\{\m}\hat{p}^{\n\}}\)\frac{1}{p}-\(\g\cdot\hat{p}\hat{q}^\m\hat{q}^\n+2\g\cdot\hat{q}\hat{p}^{\{\m}\hat{q}^{\n\}}\)\frac{1}{p}-2\g\cdot\hat{p}\hat{p}^{\{\m}u^{\n\}}\nonumber\\
    -&2\g\cdot\hat{l}\hat{l}^{\{\m}P^{\n\}}\frac{\ln\frac{2p}{q}}{p^2}+\g\cdot\hat{p}P^\m P^\n\frac{1}{p^3} \big].
\end{align}

Finally we sum over \eqref{I_exp}, \eqref{II_exp}, \eqref{III_exp} and \eqref{IV_exp} and note $4\p a=\frac{e^2T^2}{8}=m_f^2$ to obtain
\begin{align}\label{final}
  \d\G^{\m\n}=m_f^2\big[-\g\cdot\hat{p}P^\m P^\n\frac{\ln\frac{2p}{q}}{p^3}-\g\cdot\hat{l}P^{\{\m}\hat{l}^{\n\}}\frac{\ln\frac{2p}{q}}{p^2}+\g\cdot\hat{p}\(2u^\m u^\n+u^{\{\m}\hat{p}^{\n\}}+\hat{p}^\m\hat{p}^\n\)\frac{1}{p}+2\g\cdot\hat{l}\hat{l}^{\{\m}\hat{p}^{\n\}}\big].
\end{align}

\begin{figure}[t]
\includegraphics[width=0.4\textwidth]{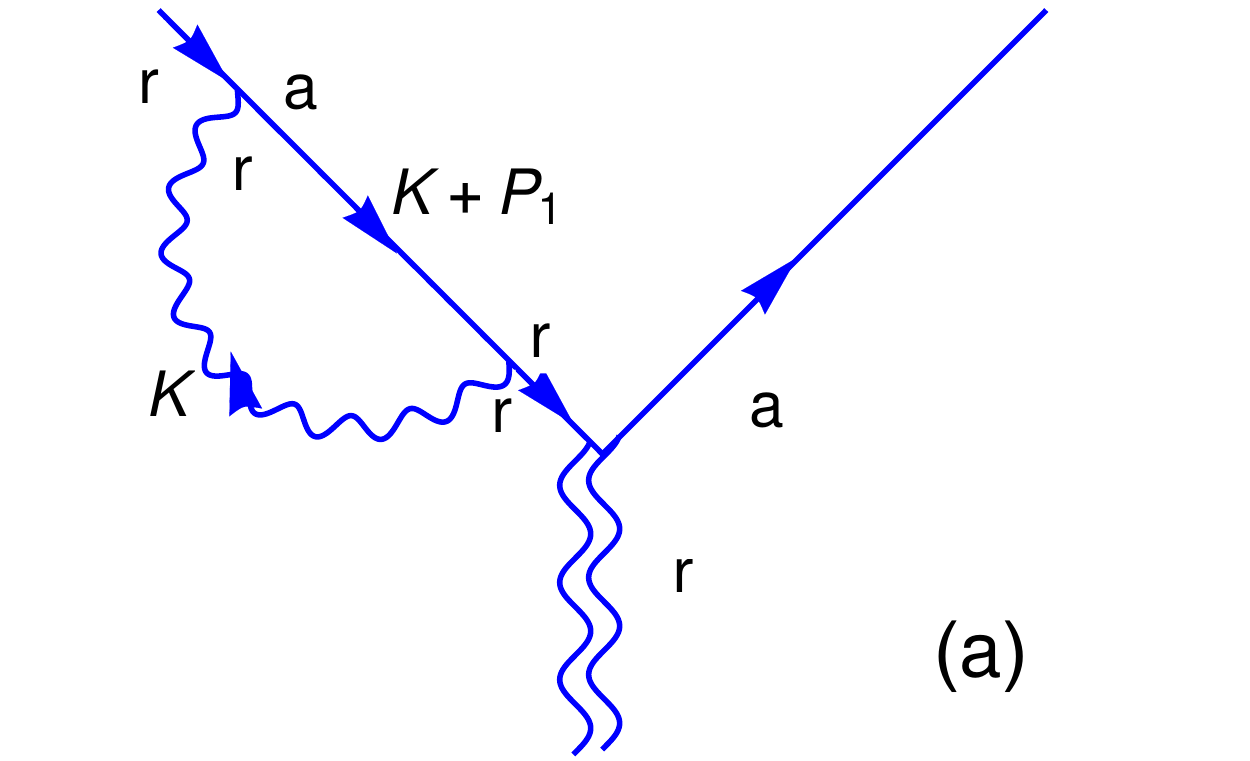}
\includegraphics[width=0.4\textwidth]{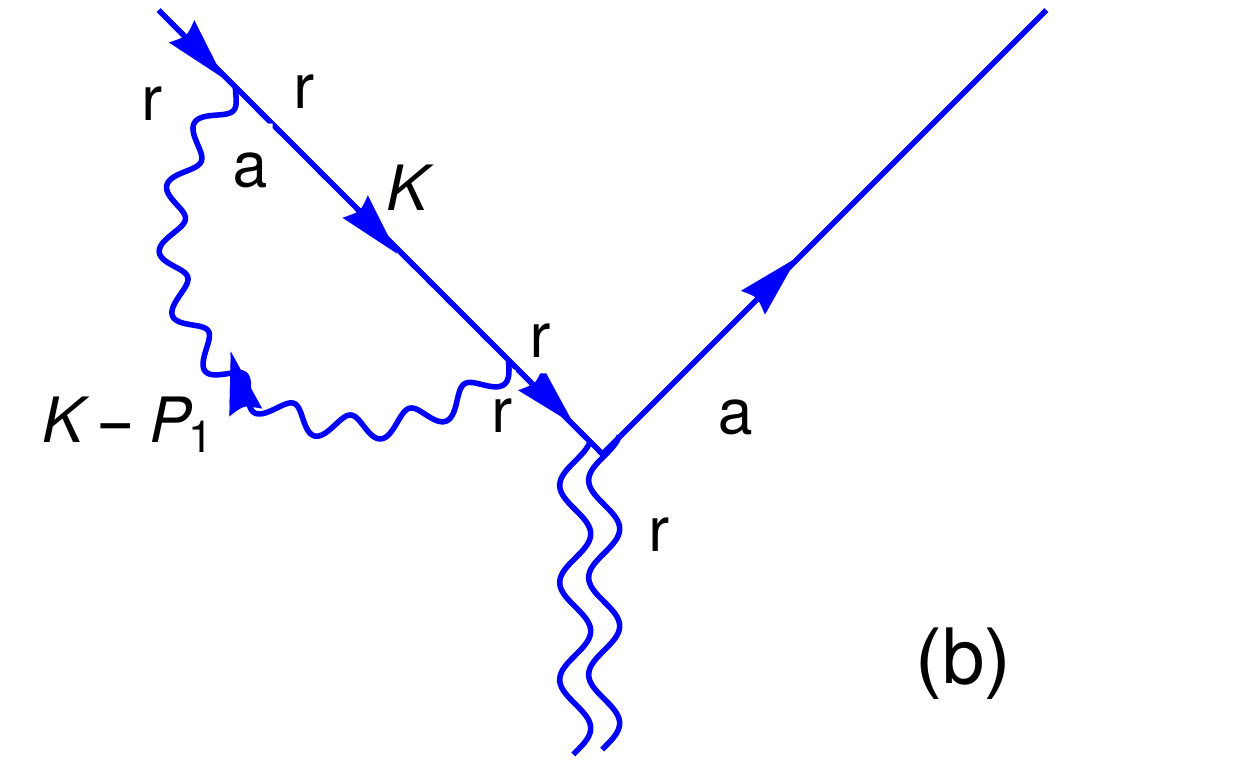}
\caption{(color online) Diagrams for radiative correction to external fermion leg, with arrows indicating direction of momenta. Similar diagrams for correction to the other external leg not shown.}\label{fig_diag4}
\end{figure}

We still need to consider fermion self-energy diagram Fig.~\ref{fig_diag4}. Similar to vacuum situation, the effect of self-energy changes the on-shell condition of fermion, and introduces the field strength renormalization. In the medium, the resummed fermion propagator becomes \cite{Bellac:2011kqa}
\begin{align}\label{resum_prop}
  S^{ra}(P)=\frac{i}{2}\D_+(P)\(\g^0-\g\cdot\hat{p}\)+\frac{i}{2}\D_-(P)\(\g^0+\g\cdot\hat{p}\),
\end{align}
with $\D_\pm(P)=\(p_0\mp p-\frac{m_f^2}{2p}\big[\(1\mp\frac{p_0}{p}\)\ln\frac{p_0+p}{p_0-p}\pm 2\big]\)^{-1}$. Here the pole of $\D_+$ corresponds to medium modified fermion energy, while counterpart of $\D_-$ gives the plasmino mode present in medium only. Since we are concerned with medium correction to fermion spin-vorticity coupling in vacuum, we will only consider $\D_+$. Here the self-energy correction is formally the same as in vacuum, coming entirely from the field strength renormalization
\begin{align}\label{self}
  \d\G^{\m\n}=\d Z_+\g^{\{\m}P^{\n\}}.
\end{align}
In the limit $p\gg m_f$, the field strength renormalization has a simple expression \cite{Bellac:2011kqa}
\begin{align}\label{Z}
  \d Z_+=\frac{m_f^2}{2p^2}\(1-\ln\frac{2p^2}{m_f^2}\).
\end{align}
Note that we still have $p\ll T$. The sum of \eqref{final} and \eqref{self} gives the medium correction to the GFF.

\subsection{Result and discussions}\label{subsec_disc}

Now we discuss result of the GFF. We consider corrections to spin-vorticity coupling from \eqref{final} and \eqref{self}. To ease the discussions, we follow the treatment of the previous subsection to point $\vec{p}$ and $\vec{q}$ along $z$ and $y$ respectively. For massless fermion, the spin direction points approximately along $z$, thus we need to introduce $0x$ component of metric perturbation to induce vorticity along $z$. Possible spin-vorticity coupling comes from the interaction vertex $h_{\m\n}\d\G^{\m\n}$, with only the $0x$ component of $\d\G^{\m\n}$ contributing. We first consider contribution from \eqref{self}. It differs from the vacuum GFF only by a constant factor $\d Z_+$. From the second term on the RHS of \eqref{gravFF0}, we obtain its contribution to scattering amplitude as
\begin{align}\label{M_self}
  i{\cal M}&=i\bar{u}(P_2)\frac{\s^3}{2}q\frac{\tilde{v}}{2}u(P_1)\d Z_+\nonumber\\
  &\simeq2p\frac{i\tilde{\o}}{2}\frac{m_f^2}{2p^2}\(1-\ln\frac{2p^2}{m_f^2}\),
\end{align}
with $\tilde{\o}=-\frac{i}{2}q\tilde{v}$ and the prefactor $2p$ coming from the normalization of relativistic spinor. On the other hand, the contribution to scattering amplitude from (the second term of) \eqref{final} reads
\begin{align}\label{M_final}
  i{\cal M}&=i\bar{u}(P_2)(-\s^1)u(P_1)\frac{\tilde{v}}{2}m_f^2\frac{-\ln\frac{2p}{q}}{p}\nonumber\\
  &\simeq2p\frac{i\tilde{\o}}{2}\frac{m_f^2}{p^2}\ln\frac{2p}{q},
\end{align}
here we have used \eqref{usigma}. We stress that although \eqref{M_self} and \eqref{M_final} give similar contributions, their physical explanations are different. To see that, we formally expand the scattering amplitude: $i{\cal M}\sim\langle P_2|-i\int dt H|P_1\rangle$, with $H\sim\bar\ps\d\G^{\m\n}\ps h_{\m\n}$. Note that we have imposed the constraint $q_0=0$ in the calculation of $\d\G^{\m\n}$, which requires us to take time independent $h_{\m\n}$ when applying to scattering problem. Only this way $\int dt$ can gives the factor $\d(q_0)$. The remaining part is the transition matrix element of energy operator. Below we clarify that only \eqref{M_self} corresponds to potential: in \eqref{M_self} we have ignored the difference between the initial and final states, so that $\bar{u}(P_2)\frac{\s^3}{2}u(P_1)$ can be viewed as the spin of the same state. Since the vorticity is time independent, \eqref{M_self} can be naturally interpreted as potential, i.e. medium correction to the spin-vorticity coupling. However the $q$ in \eqref{M_final} comes from $\bar{u}(P_2)\s^1u(P_1)$. Here ignoring the difference between the initial and final states would lead to a vanishing result. Therefore \eqref{M_final} cannot be explained as energy of the same state, but transition matrix element between initial and final states. In fact, $\langle P_2|\d\G^{\m\n}|P_1\rangle$ is generically matrix element of EMT between initial and final states. Only particular components adopt simple potential interpretation.

Although different in interpretation, both of them contribute to ACVE. We explain briefly as follows: $\d\G^{\m\n}$ is effective vertex correction including graviton-fermion vertex correction and fermion self-energy correction. It gives the following form of radiative correction to axial current
\begin{align}
  J_5^\r\sim\int_P\tr\[\g^5\g^\r{\slashed P}_1\d\G^{\m\n}{\slashed P}_2\]\frac{h_{\m\n}}{2}.
\end{align}
With our choices of momenta and metric perturbation, we easily find that both \eqref{self} and \eqref{final} contribute to axial current along $z$:
\begin{align}\label{trace_identity}
  &\tr\[\g^5\g^3{\slashed P}_1\g^{\{\m}P^{\n\}}{\slashed P}_2\]h_{\m\n}=\tr\[\g^5\g^3{\slashed P}_1(-\g\cdot\hat{l}P^{\{\m}\hat{l}^{\n\}}){\slashed P}_2\]h_{\m\n}\propto ip^2qv\sim p^2\o.
\end{align}

Finally we analyze the infrared divergence of the logarithmic term as $q\to0$. This divergence occurs in $II$ and $IV$. If we further restrict ourselves to the divergent terms coupling to vorticity, i.e. terms with $\m\n=0x$. It comes from $IV$ only, which can be traced back to the case $\l\m\n=xxz$ in \eqref{IV_integral2}. We discuss this case separately in appendix, which finally reduces to the integral in \eqref{xx_integral}. In the limit $q\to0$, we find the integrand contains non-integrable collinear divergence\footnote{In fact this collinear divergence differs slightly from the ones canceled in the earlier discussions. Here the collinearity of $K$ and $P_{1,2}$ occurs simultaneously with $K\cdot Q\to0$.}. $q\ne0$ turns the non-integrable divergence into a logarithmic one. The logarithmic divergence can be cut off by screening effect of the medium: in the medium both fermion and photon gain thermal masses such that their momenta are no longer light-like, which is sufficient to cut off the divergence. When $q$ is much less than the thermal masses, it is not difficult to imagine that thermal masses should replace $q$ as the infrared cutoff. Based on this argument, we expect partial cancellation between \eqref{M_self} and \eqref{M_final}. Below we consider screening effect of the medium explicitly.

We first write down the logarithmic divergent structures $IV$
\begin{align}
  -4K^\m K^\n{\slashed K}\frac{1}{2K\cdot P_1}\frac{Q^2}{(2K\cdot Q)^2}\(f(k_0)+\tilde{f}(k_0)\)+(P_1\to P_2).
\end{align}
We will see the screening effect renders the result infrared safe in the limit $q\to0$, for which we replace $P_1$ and $P_2$ by $P$ below. In the above expression, the distribution functions $\tilde{f}(k_0)$ and $f(k_0)$ come from Fig.~\ref{fig_diag2} and Fig.~\ref{fig_diag3}. The factors leading to the infrared divergences in Fig.~\ref{fig_diag2} and Fig.~\ref{fig_diag3} correspond to fermion and photon propagators respectively. With screening effect of the medium, we need to substitute them with the corresponding resummed propagators. The resummed fermion propagator is already given in \eqref{resum_prop}. We need the case $P\to L=K-P$. Since $K\gg P$, we can expand $\D_\pm$ to obtain
\begin{align}
  &\D_+\simeq\(-p+p\cdot\hat{k}-\frac{m_f^2}{k}\)^{-1},\nonumber\\
  &\D_-\simeq(2k)^{-1}.
\end{align}
\begin{align}
  &\D_+\simeq\(-p+p\cdot\hat{k}-\frac{m_f^2}{k}\)^{-1},\nonumber\\
  &\D_-\simeq(2k)^{-1}.
\end{align}
Noting that $\D_-\ll \D_+$, we can keep only the $\D_+$ component, and use $\hat{l}\simeq\hat{k}$ to approximate the propagator as
\begin{align}
  S^{ra}\simeq\frac{i}{2(-p+p\cdot\hat{k}-\frac{m_f^2}{k})}\(\g^0-\g\cdot\hat{k}\)=\frac{i{\slashed K}}{-2(K\cdot P+m_f^2)}.
\end{align}
This amounts to the following substitution to the collinear factor from the fermion propagator:
\begin{align}\label{replace_fermion}
  \frac{1}{2K\cdot P}\to\frac{1}{2K\cdot P+m_f^2}.
\end{align}
For the case in which the collinear factor is from the photon propagator, we choose the resummed photon propagator in Coulomb gauge \cite{Bellac:2011kqa}.
\begin{align}
  D_{\m\n}^{ra}(L)=\frac{i}{L^2-\P_T^R}P_{\m\n}^T(L)+\frac{i}{l^2-\P_L^R}u_\m u_\n,
\end{align}
with
\begin{align}\label{Pi}
  &\P_T^R=m_\g^2\frac{l_0}{l}\big[\(1-\frac{l_0^2}{l^2}\)Q_0\(\frac{l_0}{l}\)+\frac{l_0}{l}\big],\nonumber\\
  &\P_L^R=-\frac{-l^2}{L^2}2(m_\g^2-\P_T^R).
\end{align}
Noting $L=K-P$ and $K\gg P$, we easily find the longitudinal component of the propagator can be ignored, and the remaining transverse component can be approximated as
\begin{align}
  D_{\m\n}^{ra}\simeq\frac{i}{-2K\cdot P-m_\g^2}P_{\m\n}^T(K).
\end{align}
Here apart from the substitution of the collinear factor
\begin{align}\label{replace_photon}
  \frac{1}{2K\cdot P}\to\frac{1}{2K\cdot P+m_\g^2},
\end{align}
there is also change of the polarization tensor $g_{\m\n}\to-P_{\m\n}^T$. In fact we can easily show using the property of transverse projection operator $P_{\m\n}^T(K)$ that the change of polarization tensor does not change the final Dirac structure. Therefore the modification from the screening effect is only the substitutions \eqref{replace_fermion} and \eqref{replace_photon}, in which $P$ can be either $P_1$ or $P_2$.

The substitutions above can be implemented by letting $a=(p^2+\frac{q^2}{4})^{1/2}-p\cos\th+\frac{m^2}{k}$ in \eqref{xx_integral}. Depending on the types of the corresponding propagators, $m^2$ can be either $m_f^2$ or $m_\g^2$. When $q\ll\frac{m^2}{k}\sim e^2T$, we can make the following approximation $a\simeq p-p\cos\th+\frac{m^2}{k}$ and ignore $b$ in \eqref{xx_integral} to arrive at the following result
\begin{align}\label{xx_integral_reg}
  \int d\cos\th\sin^2\th\frac{-2\pi}{ac^2}=-\frac{2\p\ln\(1+\frac{2pk}{m^2}\)}{pq^2}\simeq -\frac{2\p\ln\frac{2pk}{m^2}}{pq^2}.
\end{align}
The effect of regularization by the thermal masses amounts to the following substitution in the coefficient of $-\g\cdot\hat{l}P^{\{\m}\hat{l}^{\n\}}\frac{1}{p^2}$ in \eqref{final}
\begin{align}\label{replace_coeff}
  &4\p e^2\int\frac{kdk}{(2\p)^2}\ln\frac{2p}{q}\(\tilde{f}(k_0)+f(k_0)\)\to 2\p e^2\int\frac{kdk}{(2\p)^2}\(\ln\frac{2pk}{m_\g^2}\tilde{f}(k_0)+\ln\frac{2pk}{m_f^2}f(k_0)\),\nonumber\\
\Rightarrow  &m_f^2\ln\frac{2p}{q}\to \frac{m_f^2}{2}\(\frac{1}{3}\ln\frac{2p}{m_\g^2}+\frac{2}{3}\ln\frac{2p}{m_f^2}+1-12\ln A+\frac{1}{3}\ln(16\p^3T^3)\).
\end{align}
Here $A\simeq 1.282$ is the Glaisher constant.

In the analysis earlier, we have seen that although the Dirac structure $-\g\cdot\hat{l}P^{\{\m}\hat{l}^{\n\}}$ cannot be interpreted as correction to spin-vorticity coupling, it still contributes to ACVE. In particular, the identity in \eqref{trace_identity} indicates that its contribution to the axial current equals to the counterpart from the Dirac structure $\g^{\{\m}P^{\n\}}$. Thus we can sum over the coefficients of the two structures directly.
\begin{align}\label{CVE_total}
  &\frac{m_f^2}{2p^2}\(1-\ln\frac{2p^2}{m_f^2}\)+\frac{m_f^2}{2p^2}\(\frac{1}{3}\ln\frac{2p}{m_\g^2}+\frac{2}{3}\ln\frac{2p}{m_f^2}+1-12\ln A+\frac{1}{3}\ln(16\p^3T^3)\)\nonumber\\
  =&\frac{m_f^2}{p^2}\(1-6\ln A+\frac{1}{6}\ln(16\pi^3)\)+\frac{m_f^2}{2p^2}\ln\frac{T}{p}+\frac{m_f^2}{6p^2}\ln\frac{m_f^2}{m_\g^2}.
\end{align}
We point out two interesting features of the result above: i. the thermal masses of fermion and photon appears as ratio in the logarithm. Quoting the explicit results $m_f^2=\frac{1}{8}e^2T^2$, $m_\g^2=\frac{1}{6}e^2T^2$, we find no logarithmic enhancement like $\ln e^{-1}$ in the result; ii. because $p\ll T$, the dominant contribution in the above is the second term, which is positive. In fact, all three terms in \eqref{CVE_total} are separately positive. These features are in qualitative agreement with the result in \cite{Hou:2012xg}. Although we have used the kinematic restriction $p\ll T$, the regime does not give dominant contribution to axial current. Nevertheless, the agreement with \cite{Hou:2012xg} implies that the GFF we have obtained might reflect qualitative features of fermion scattering with graviton at $p\sim T$.


\section{Summary and outlook}\label{sec_outlook}

In this paper, we have studied the medium correction to the GFF. We first generalize the vacuum GFF to massless case. By introducing suitable metric perturbation to mimic fluid vorticity, we find the vacuum GFF can describe the well-known spin-vorticity coupling. By equivalence principle, radiative correction in vacuum cannot renormalize the spin-vorticity coupling.

The equivalence principle based on Lorentz invariance is violated in a medium. Using quantum electrodynamic plasma as an example, we have studied the medium correction to the GFF, and discussed possible medium correction to spin-vorticity coupling. In the HTL approximation, we find only two structures contributing to fermion scattering in vorticity field. One structure comes from fermion self-energy correction, which can be interpreted as medium correction to spin-vorticity coupling. Our result points to suppression of spin-vorticity coupling in medium. The other structure comes from graviton-fermion vertex correction, which does not adopt potential interpretation, but corresponds to transition matrix element between initial and final states. This contribution is infrared divergent as the momentum exchanges tends to zero. We have obtained infrared safe result after introducing screening effect of the medium. Our analysis indicates that both structures contribute to ACVE. Combining the two contributions, we find an enhanced net axial current from the radiative correction, which is in qualitative agreement with the known result.

Given that the diagrams considered in this paper have one-to-one correspondence in quantum chromodynamics, we believe the conclusion above applies to the latter as well. For spin polarization effect in local equilibrium, only the first structure corresponding to spin-vorticity coupling plays a role. Thus we only need to consider contribution from the fermion self-energy. The only change needed is the expression for thermal mass, for which $m_f^2=\frac{1}{8}g^2T^2C_F$ with $C_F=\frac{4}{3}$. With applications to heavy ion collisions phenomenology in mind, we further take $\a_s=0.3$, $T=350\text{MeV}$ and $p=1\text{GeV}$. Note that the HTL does not apply, but given the suppression comes from fermion self-energy renormalization, one expect medium suppression to spin-vorticity coupling might be generic. We obtain about $9\%$ suppression in spin-vorticity coupling based on \eqref{self}. This implies the so far ignored radiative correction might have an appreciable effect in spin polarization phenomenon in heavy ion collisions.

The medium correction to the GFF obtained in this paper can be used to study the couplings of spin to all fluid gradient. We have considered only spin-vorticity coupling and kept only the real part of the form factors. Because the spin-vorticity coupling has simple potential form, so the real part is sufficient to describe their interaction. The couplings of spin and other form of fluid gradient such as shear might lead to spin dissipation. The dissipative effect can be described by imaginary part of the GFF. The spin dissipation can be an indispensable ingredient in spin polarization phenomenon in heavy ion collisions.

This paper focuses on the GFF of fermion. Similar discussions apply to electromagnetic form factors. An interesting question is how does medium modifies the spin-magnetic coupling in chiral limit? This can affect thermodynamics of fermions in magnetic field \cite{Zhang:2020ben,Fang:2021ndj} as well as the chiral magnetic effect, see \cite{acta_cme} for a recent review. Another possible extension is to non-relativistic Weyl fermion, the corresponding electromagnetic form factors can be used to study radiative corrections to non-relativistic chiral kinetic theory \cite{Gao:2022gqr}.

Finally the method used in this paper can also be generalized to composite particles such as vector meson. Recent experiments have revealed spin alignment of vector mesons, see \cite{acta_ex,acta_th,Wang:2023fvy} for related discussions. The GFF for vector meson can provide a new way to describe coupling of vector meson spin with fluid gradient. We will report progress in future works.

\section*{Acknowledgments}
S.L. is indebted to Defu Hou for insightful discussions. This work is in part supported by NSFC under Grant Nos 12075328, 11735007.

\appendix

\section{Regularization of integrals in $IV$}\label{sec_app}

We need to calculate the following types of integrals
\begin{align}\label{IV_reg}
  &\(K^\l-\frac{K\cdot u}{P_1\cdot u}P_1^\l\)\(K^\m-\frac{K\cdot u}{P_1\cdot u}P_1^\m\)\(K^\n-\frac{K\cdot u}{P_1\cdot u}P_1^\n\)\frac{1}{2K\cdot P_1}\frac{Q^2}{(2K\cdot Q)^2}\(f(k_0)+\tilde{f}(k_0)\)\nonumber\\
  &\(K^\l-\frac{K\cdot u}{P_1\cdot u}P_1^\l\)\big[\(K^\m-\frac{K\cdot u}{P_1\cdot u}P_1^\m\)\frac{K\cdot u}{P_1\cdot u}P_1^\n+(\m\leftrightarrow\n)\big]\frac{1}{2K\cdot P_1}\frac{Q^2}{(2K\cdot Q)^2}\(f(k_0)+\tilde{f}(k_0)\)\nonumber\\
  &\(K^\l-\frac{K\cdot u}{P_1\cdot u}P_1^\l\)\(\frac{K\cdot u}{P_1\cdot u}\)^2P_1^\m P_1^\n\frac{1}{2K\cdot P_1}\frac{Q^2}{(2K\cdot Q)^2}\(f(k_0)+\tilde{f}(k_0)\)
\end{align}
Here the divergence from the denominator $(2K\cdot Q)^2$ appears unavoidable. To solve this problem, we need to restore $i\e$. Note that the denominators above come from expansion of propagators $D_{ra}(K-Q)$ and $D_{ar}(K+Q)$. Thus we need to make the following substitutions in the expansion $-2K\cdot Q\to -2K\cdot Q-i\sgn(k_0)\e$, $2K\cdot Q\to 2K\cdot Q+i\sgn(k_0)\e$. Given that $k_0=\pm k$ still gives identical contribution with the regularization, we take $k_0=k$ and replace the denominator by $(2K\cdot Q+i\e)^2$. To simplify notations, we define
\begin{align}\label{abc}
  a=\(p^2+\frac{q^2}{4}\)^{1/2}-p\cos\th,\quad
  b=\frac{q}{2}\sin\th,\quad
  c=\frac{q}{2}\sin\th.
\end{align}
The indices of the first term in \eqref{IV_reg} can take $\l\m\n=zzz,\;zxx,\;yxx,\;zyy,\;zzy$, the indices of the second term can take $\l\m=zz,\;xx,\;zy$ and the index of the third term takes $\l=z$. In the limit $q\to0$, the remaining indices can only take $0$ or $z$. Below we illustrate the angular integrations using $\l\m\n=zzz$ for the first term as an example. The corresponding angular dependent part of the integral can be written as
\begin{align}
  \int d\cos\th d\vj\(\cos\th-\frac{p}{(p^2+\frac{q^2}{4})^{1/2}}\)^3\frac{1}{4(a+b\sin\vj)(c\sin\vj+i\e)^2},
\end{align}
The $\vj$ integration can be performed first to obtain
\begin{align}\label{phi_int}
  \int d\cos\th\(\cos\th-\frac{p}{(p^2+\frac{q^2}{4})^{1/2}}\)^3\frac{2\p\(\frac{b^2}{a^2-b^2}-\frac{\e ac^2}{(c^2+\e^2)^{3/2}}+\frac{ibc(c^2+2\e^2)}{(c^2+\e^2)^{3/2}}\)}{4(ac-i\e b)^2}.
\end{align}
We take the limit $\e\to0$, in which the denominator of \eqref{phi_int} simplifies to $(ac-i\e b)^2\to (ac)^2$. Since we are concerned with the spin-vorticity coupling corresponding to the real part, we drop the last purely imaginary term in the numerator. The second term tends to zero superficially, but actually not. Because $c=\frac{q}{2}\sin\th$, the integration domain $\sin\th\sim\e$ can give rise to significant contribution thus needs to be kept. Performing the integrations over $\cos\th$ for the first and second terms, we obtain
\begin{align}\label{th_int}
  &\int d\cos\th\(\cos\th-\frac{p}{(p^2+\frac{q^2}{4})^{1/2}}\)^3\frac{2\p\(\frac{b^2}{a^2-b^2}\)}{4(ac)^2}
  \simeq-\frac{\p}{p^3},\nonumber\\
  &\int d\cos\th\(\cos\th-\frac{p}{(p^2+\frac{q^2}{4})^{1/2}}\)^3\frac{2\p\(-\frac{\e ac^2}{(c^2+\e^2)^{3/2}}\)}{4(ac)^2}
  \simeq\frac{8\p}{pq^2},
\end{align}
where in the result of the second term we have taken the limit $\e\to0$. $\simeq$ indicates that we have kept dominant contribution in the limit $q\to0$. Comparing the two terms we find the contribution from the first term can be ignored in the limit $q\to0$.

To ease the treatment of infrared divergence, we discuss separately the case with $\l\m=xx$ for the second term. The corresponding integral can be written as
\begin{align}
  \int d\cos\th d\vj\frac{\cos^2\vj\sin^2\th}{4(a+b\sin\vj)(c\sin\vj+i\e)^2}.
\end{align}
The integration of $\vj$ gives
\begin{align}
  \int d\cos\th\sin^2\th\frac{2\p\(ibc+a\e-\sqrt{(a^2-b^2)(c^2+\e^2)}\)}{(ac-ib\e)^2\sqrt{c^2+\e^2}}.
\end{align}
We still convert the denominator as $(ac-ib\e)^2\to (ac)^2$. In the numerator, the first term is purely imaginary, and the integrand of the second term is finite only when $\sin\th\sim\e$. But at the same time the integration domain $d\cos\th\sim\e$, therefore the result tends to zero as $\e\to0$. We are left with the third term
\begin{align}\label{xx_integral}
  \int d\cos\th\sin^2\th\frac{-2\pi\sqrt{a^2-b^2}}{4(ac)^2}\simeq-\frac{4\p\ln\frac{2p}{q}}{pq^2}.
\end{align}
Other cases can be obtained similarly. We do not elaborate here.

\end{CJK}

\end{document}